\documentstyle[prl,tighten,aps,epsfig]{revtex}

\def\beqa{\begin{eqnarray}}
\def\eeqa{\end{eqnarray}}
\def\beq{\begin{equation}}
\def\eeq{\end{equation}}

\def\vol{\int d^4x\,\sqrt{-g}}

\def\half{\frac{1}{2}}
\def\gu{g^{\mu\nu}}
\def\gd{g_{\mu\nu}}

\def\dmu{_{\mu}}
\def\dnu{_{\nu}}
\def\umunu{^{\mu\nu}}
\def\dmunu{_{\mu\nu}}

\def\da{_{\alpha}}

\def\uab{^{\alpha\beta}}
\def\dab{_{\alpha\beta}}
\def\dabgd{_{\alpha\beta\gamma\delta}}

\def\ddemunu{_{;\mu\nu}}
\def\udemunu{^{;\mu\nu}}
\def\ddemu{_{;\mu}}  \def\udemu{^{;\mu}}
\def\ddenu{_{;\nu}}  \def\udenu{^{;\nu}}
\def\ddea{_{;\alpha}}  \def\udea{^{;\alpha}}
\def\ddeb{_{;\beta}}

\def\pa{\partial}

\let\lam=\lambda
\let\Lam=\Lambda

\let\gam=\gamma
\let\alp=\alpha
\let\sig=\sigma
\let\lb=\label
\renewcommand{\epsilon}{\varepsilon}
\let\no=\nonumber
\let\noin=\noindent
\let\in=\indent

\def\bet{\begin{tabular}}
\def\eet{\end{tabular}}
\def\bef{\begin{figure}}
\def\eef{\end{figure}}
\def\beqa{\begin{eqnarray}}
\def\eeqa{\end{eqnarray}}
\def\beq{\begin{equation}}
\def\eeq{\end{equation}}
\def\vol{\int d^4x\,\sqrt{-g}}

\def\half{\frac{1}{2}}
\def\ra{\rightarrow}
\def\Gu{G^{\mu\nu}}
\def\Gd{G_{\mu\nu}}
\def\gu{g^{\mu\nu}}
\def\gd{g_{\mu\nu}}
\def\hu{h^{\mu\nu}}
\def\hd{h_{\mu\nu}}

\def\dmu{_{\mu}}
\def\dnu{_{\nu}}
\def\umunu{^{\mu\nu}}
\def\dmunu{_{\mu\nu}}

\def\da{_{\alpha}}

\def\uab{^{\alpha\beta}}
\def\dab{_{\alpha\beta}}
\def\dabgd{_{\alpha\beta\gamma\delta}}

\def\ddemunu{_{;\mu\nu}}
\def\udemunu{^{;\mu\nu}}
\def\ddemu{_{;\mu}}
\def\udemu{^{;\mu}}
\def\ddenu{_{;\nu}}
\def\udenu{^{;\nu}}
\def\ddea{_{;\alpha}}
\def\udea{^{;\alpha}}
\def\ddeb{_{;\beta}}

\def\pa{\partial}
\def\etal{{\it et al.}}
\def\ie{{\it i.e. }}
\def\eg{{\it e.g. }}

\def\p{\phi}
\def\pv{\varphi}
\def\v{V(\phi)}
\def\vp{V'(\phi)}
\def\l{\cal L}
\def\g{\cal G}
\def\fo{\cal F}
\def\s{\cal S}

\let\lam=\lambda
\let\Lam=\Lambda

\let\gam=\gamma
\let\alp=\alpha
\let\sig=\sigma
\let\lb=\label
\renewcommand{\epsilon}{\varepsilon}
\let\no=\nonumber
\let\noin=\noindent
\let\in=\indent

\let\lam=\lambda  \let\Lam=\Lambda

\let\gam=\gamma
\let\alp=\alpha
\let\sig=\sigma

\let\lb=\label
\renewcommand{\epsilon}{\varepsilon}
\let\no=\nonumber

\def\pr{{\it Phys. Rev.}\ }
\def\prl{{\it Phys. Rev. Lett.}\ }
\def\pl{{\it Phys. Lett.}\ }
\def\np{{\it Nucl. Phys.}\ }

\def\ijmp{{\it Int. Journ. Mod. Phys.}\ }

\def\cqg{{\it Class. Quantum Grav.}\ }

\def\grg{{\it Gen. Relativ. Grav.}\ }

\def\apj{{\it Ap. J.}\ }

\def\mnras{{\it Mon. Not. R. Ast. Soc.}\ }

\def\araa{{\it Ann. Rev. Astr. Ap.}\ }

\def\rmp{{\it Rev. Mod. Phys.}\ }
\def\etal{{\it et al.}}
\def\ie{{\it i.e. }}
\def\eg{{\it e.g. }}

\def\p{\phi}
\def\pv{\varphi}
\def\v{V(\phi)}
\def\vp{V'(\phi)}
\def\l{\cal L}

\begin{document}
\begin{titlepage}
        \title{Newtonian limit of Extended Theories of Gravity}
\author{Salvatore Capozziello\thanks{e-mail: capozziello@sa.infn.it} \\
{\em Dipartimento di Fisica "E.R. Caianiello"} \\
 {\em Universit\'a di Salerno, 84081 Baronissi (Sa), Italy.} \\
 {\em Istituto Nazionale di Fisica Nucleare, Sez. di Napoli, Italy.} \\ }
\date{\today}
\maketitle

\begin{abstract}
Newtonian limit of Extended Theories of Gravity (in particular,
higher--order and scalar--tensor theories) is theoretically
discussed taking into account recent observational and
experimental results.  Extended Theories of Gravity have become a
sort of paradigm in the study of gravitational interaction since
several motivations push for enlarging the traditional scheme of
Einstein General Relativity. Such issues come, essentially, from
cosmology and quantum field theory. In the first case, it is well
known that higher--derivative theories  and scalar--tensor
theories  furnish inflationary cosmological solutions capable, in
principle, of solving the shortcomings of The Standard
Cosmological Model. Besides, they have relevant features also from
the quantum cosmology viewpoint. In the second case, every
unification scheme as Superstrings, Supergravity or Grand Unified
Theories, takes into account effective actions where nonminimal
couplings to the geometry or higher--order terms in the curvature
invariants come out. Such contributions are due to one--loop or
higher--loop corrections in the high--curvature regimes near the
full (not yet available) quantum gravity regime. In the
weak--limit approximation, all these classes of theories should be
expected to reproduce the Einstein General Relativity which, in
any case, is experimentally tested only in this limit. This fact
is matter of debate since several relativistic theories
 do not reproduce the Einstein results in the Newtonian approximation.
 For example, some authors claim for explaining the flat rotation
curves of galaxies by  Yukawa--like corrections to the Newtonian
potential arising in this context. Others  have shown that a
conformal theory of gravity  could explain the missing matter
problem. Moreover, indications of an apparent, anomalous,
long--range acceleration revealed from the data analysis of
Pioneer 10/11, Galileo, and Ulysses spacecrafts could be framed in
a general theoretical scheme by taking into account Yukawa--like
or higher--order corrections to the Newtonian potential. Such data
will be very likely capable of confirming or ruling out the
physical consistency of Extended Theories of Gravity.

\end{abstract}
\vspace{10.mm}  Keyword(s): Extended theories of gravity,
weak-field approximation, Newtonian limit. \vfill
\end{titlepage}

\section{\bf Introduction}
Einstein's General Relativity (GR) can be considered one of the
major scientific achievement of last century. For the first time,
a comprehensive theory of spacetime, gravity and matter has been
formulated giving rise to a new conception of the Universe.
However, in the last thirty years, several shortcomings came out
in the Einstein scheme and people  began to investigate if GR is
the only and fundamental theory  capable of explaining completely
gravitational interaction. Such issues come, essentially, from
cosmology and quantum field theory. In the first case, the
presence of the Big Bang singularity, flatness and horizon
problems \cite{guth} led to the statement that Standard
Cosmological Model \cite{weinberg}, based on GR and Standard Model
of particle physics, is inadequate to describe the Universe in
extreme regimes. On the other hand, GR is a {\it classical} theory
which does not work as a fundamental theory, if one wants to
achieve a full quantum description of spacetime (and then of
gravity). Due to this facts and, first of all, to the lack of a
definitive quantum gravity theory, alternative theories of gravity
have been pursued in order to attempt, at least, a semi-classical
scheme where GR and its positive results could be recovered.

One of the most fruitful approach has been that of {\it Extended
Theories of Gravity} (ETG)  which have become a sort of paradigm
in the study of gravitational interaction based on the enlargement
and the correction  of the traditional Einstein scheme. The
paradigm consists in adding higher-order curvature invariants and
minimally or non-minimally coupled scalar fields into dynamics
which come out from quantum terms in the effective action of
gravity. Other motivations come from the issue of a complete
recovering of Mach principle \cite{brans} which leads to assume a
varying gravitational coupling. All these approaches are not the
"{\it full quantum gravity}" but are needed as working schemes
toward it. In  any case, they are going to furnish consistent and
physically reliable results.

For example, it is well known that higher--derivative theories
\cite{starobinsky} and scalar--tensor theories \cite{la} give rise
to inflationary cosmological solutions capable, in principle, of
solving the shortcomings of Standard Cosmological Model. Besides,
they have relevant features also from the quantum cosmology point
of view since they give interesting solutions to the problem of
initial, at least in the restricted context of minisuperspaces
\cite{vilenkin}.

Furthermore, every unification scheme as Superstrings,
Supergravity or Grand Unified Theories, takes into account
effective actions where nonminimal couplings to the geometry or
higher--order terms in the curvature invariants come out. Such
contributions are due to one--loop or higher--loop corrections in
the high--curvature regimes near the full (not yet available)
quantum gravity regime \cite{odintsov}. However, in the
weak--limit approximation, all these classes of theories should be
expected to reproduce GR which, in any case, is experimentally
tested only in this limit \cite{will}.

This fact is matter of debate since several relativistic theories
(in particular ETG) {\it do not} reproduce exactly Einstein
results in the Newtonian approximation but, in some sense,
generalize them. In fact, as  it was firstly noticed by Stelle
\cite{stelle}, a $R^2$--theory gives rise to Yukawa--like
corrections to the Newtonian potential which could have
interesting physical consequences. For example, some authors claim
for explaining the flat rotation curves of galaxies by using such
terms \cite{sanders}. Others \cite{mannheim} have shown that a
conformal theory of gravity is nothing else but a fourth--order
theory containing such terms in the Newtonian limit and, by
invoking these results, it could be possible to explain {\it the
missing matter problem without dark matter}.

Besides, indications of an apparent, anomalous, long--range
acceleration revealed from the data analysis of Pioneer 10/11,
Galileo, and Ulysses spacecrafts could be framed in a general
theoretical scheme by taking into account Yukawa--like or higher
order corrections to the Newtonian potential \cite{anderson}.

In general, any relativistic theory of gravitation can yield
corrections to the Newton potential (see for example
\cite{schmidt}) which, in the post-Newtonian (PPN) formalism,
could furnish tests for the same theory \cite{will}.

Furthermore the newborn {\it gravitational lensing astronomy}
\cite{ehlers} is giving rise to additional tests of gravity over
small, large, and very large scales which very soon will provide
direct measurements for the variation of Newton coupling $G_{N}$
\cite{krauss}, the potential of galaxies, clusters of galaxies
\cite{nottale} and several other features of gravitating systems.
Such data will be very likely capable of confirming or ruling out
the physical consistency of GR or of any ETG.

In this review, we want to discuss the Newtonian limit of ETG,
trying to establish a link to GR by putting in evidence the common
features.
 In particular, we show
that, in the Newtonian limit, such theories can have features
which "generalize" the Newton potential and could explain several
observational results. For example, a quadratic correction to the
Newtonian potential strictly depends on the presence of a
scalar--field potential which acts, in the low energy limit, as a
cosmological constant.

The layout of the review is the following. In Sec.II, we discuss
the foundation of any metric theory of gravity: the Equivalence
Principle in all its formulations. After, we develop the
Parametrized Post Newtonian (PPN) formalism which constitutes the
basis to compare every theory of gravity. Moreover, we outline the
main features of GR discussing, in particular, the weak field
approximation, the linearized Einstein equations and the
gravitational radiation. Finally, a discussion of classical tests
of GR is pursued considering the current experimental limits of
the PPN parameters.

Sec.III is devoted to the ETG which are classified according to
their main features. We start with the minimal extension of GR
which is the inclusion of a cosmological constant in Einstein
equations. After, we discuss   scalar-tensor and  higher-order
theories of gravity deriving, in every case, the field equations.

In Sec.IV, the Newtonian limit of ETG is  discussed considering,
in particular, fourth-order and scalar-tensor theories. As an
application, we develop the approach for String-Dilaton gravity
deriving its weak field limit.

The discussion of the results and the conclusions are drawn in
Sec.V.

\section{The foundation of metric theories: The Equivalence Principle}
\lb{sec:principle}

Equivalence principle (EP) is the foundation of every metric
theory of gravity \cite{will}.\\ \in The first formulation of  EP
comes out from the theory of gravitation studied by Galileo and
Newton; it is called the Weak Equivalence Principle (WEP) and it
states that the ``inertial mass'' $m$ and the ``gravitational
mass'' $M$ of any object are equivalent. In Newtonian physics, the
``inertial mass'' $m$ is a coefficient which appears in the second
Newton law: $\vec{F}= m \, \vec{a}$ where $\vec{F}$ is the force
exerted on a mass $m$ with acceleration $\vec{a}$; in Special
Relativity (SR) (without gravitation) the ``inertial mass'' of a
body appears to be proportional to the rest energy of the body:
$E=m \, c^2$. Considering the Newtonian gravitational attraction,
one introduces the ``gravitational mass'' $M$: the gravitational
attraction force between two bodies of ``gravitational mass'' $M$
and $M'$ is $F= G_N M M'/r^2$ where $G_N$ is the Newtonian
gravitational constant and $r$ the distance
 between the two bodies. Various experiments \cite{urbino} demonstrate
 that $m\equiv M$.
 The present accuracy of this relation is of the order of $10^{-13}$;
 spatial projects
 are currently designed to achieve precision of $10^{-15}$
  \cite{microSCOPE} and $10^{-18}$ \cite{miniSTEP}.\\
\in The WEP statement implies that it is impossible to distinguish
between the effects of a gravitational field from those
experienced in an uniformly accelerated frame, using the simple
observation of the free-falling particles behavior. The WEP can be
formulated again in the following statement \cite{will}:

\begin{quote}
 {\sl If an uncharged test body is placed at an initial event in spacetime
 and given an initial velocity there, then its subsequent trajectory will be
 independent of its internal structure and composition}.
\end{quote}

\in A generalization of  WEP claims that SR is only locally valid.
It has been achieved by Einstein  after the formulation of  SR
theory where the concept of mass looses some of its uniqueness:
the mass is reduced to a manifestation of energy and momentum.
According to Einstein, it is impossible to distinguish between
uniform acceleration and an external gravitational field, not only
for free-falling particles but whatever is the experiment. This
equivalence principle has been called Einstein Equivalence
Principle (EEP); its main statements are the following
\cite{will}:

\begin{quote}
 \begin{itemize}
  \item[a)] {\sl WEP is valid};
  \item[b)] {\sl the outcome of any local non-gravitational test
  experiment is independent of  velocity of  free-falling apparatus};
  \item[c)] {\sl the outcome of any local non-gravitational
  test experiment is independent of where and when in
  the Universe it is performed}.
 \end{itemize}
\end{quote}

\in One defines as ``local non-gravitational experiment" an
experiment performed in a small-size \footnote{In order to avoid
the inhomogeneities.} freely falling laboratory. From the EEP, one
gets that the gravitational interaction depends on the curvature
of spacetime, \ie the postulates of any metric theory of gravity
have to be satisfied \cite{will}:

 \begin{itemize}
  \item[a)] {\sl spacetime is endowed with a metric $\gd$};
  \item[b)] {\sl the world lines of test bodies are geodesics of the metric};
  \item[c)] {\sl in local freely falling frames, called local Lorentz frames,
  the non-gravitational laws of physics are those of SR}.
 \end{itemize}

\in One of the predictions of this principle is the gravitational
red-shift, verified experimentally by Pound and Rebka in 1960
\cite{rebka}.\\
\in It is worth stressing that gravitational interactions are
specifically excluded from  WEP and  EEP. In order to classify
alternative theories of gravity, the Gravitational Weak
Equivalence Principle (GWEP) and the Strong Equivalence Principle
(SEP) has to be introduced. The SEP extends the EEP by including
all the laws of physics in its terms \cite{will}:

\begin{quote}
 \begin{itemize}
  \item[a)] {\sl WEP is valid for self-gravitating bodies as well as for test bodies (GWEP)};
  \item[b)] {\sl the outcome of any local test experiment is
  independent of the velocity of the free-falling apparatus};
  \item[c)] {\sl the outcome of any local test experiment is
  independent of where and when in the Universe it is performed}.
 \end{itemize}
\end{quote}

\in Therefore, the SEP contains the EEP, when gravitational forces
are ignored. Many authors claim that the only theory coherent with
the SEP is General Relativity (GR).

\subsection{The Parametrized Post Newtonian (PPN) limit}
\lb{sec:ppn_limit}

GR is not the only theory of gravitation and, up to now, at least
twenty-five alternative
 theories of gravity have been investigated from the 60's, considering the
 spacetime to be ``special relativistic''
 at a background level and treating gravitation as a Lorentz-invariant
 field on the background.\\
\in Two different classes of experiments have been studied: the
first ones testing the foundations of gravitation theory -- among
them the EP -- the second one testing the metric theories of
gravity where spacetime is endowed with a metric tensor and where
the EEP is valid. However, for several fundamental  reasons which
we are going to discuss,  extra fields might be necessary to
describe the gravitation, \eg scalar fields or higher-order
corrections in curvature invariants.\\
 \in Two sets of
equations can be distinguished \cite{urbino}.
 The first ones couple the gravitational fields to the
 non--gravitational contents of the Universe,
 \ie the matter distribution, the electromagnetic fields, etc...
 The second set of equations gives the evolution
 of non--gravitational fields. Within the framework of metric
  theories, these laws
 depend only on the metric: this
 is a consequence of the EEP and the so-called ``minimal
 coupling''.
  In most theories,
 including GR, the second set of equations is derived from the first
 one,
  which is the only
 fundamental one; however, in many situations, the two sets are decoupled.\\
\in The gravitational field studied in these approaches (without
cosmological considerations) is mainly due to the Sun and the
Eddington-Robertson expansion gives the corresponding metric.
Assuming spherical symmetry and a static gravitational field, one
can prove that there exists a coordinate system (called {\sl
isotropic}) such as

\beq
 d s^2 \;=\; B(r) \, dt^2 - A(r) \, r^2 \,- (dr^2 \sin^2 \theta \, d\p^2+ d\theta^2) \; .
\lb{eq:isotropic} \eeq

\noin $d t$ being the proper time between two neighboring
events.\\ \in The Newtonian gravitational field does not exceed
$G_N M_{\odot}/R_{\odot} c^2 \sim 2 \times 10^{-6}$, where $c$ is
the speed of light, $M_{\odot}$ is the mass of the Sun and
$R_{\odot}$ its radius. The metric is quasi-Minkowskian, $A(r)$
and $B(r)$ are dimensionless functions which depend only on $G_N$,
$M$, $c$ and $r$. Indeed, the only pure number that can be built
with these four quantities is $G_N M /r c^2$. The
Eddington-Robertson metric is a Taylor expansion of $A$ and $B$
which gives

\beqa
 d s^2 &=& \left( 1- 2 \, \alp \, \frac{G_N \, M}{r \, c^2} + 2 \, \beta \,
 \left( \frac{G_N \, M}{r c^2} \right)^2 + ...\right) dt^2 +\no \\
     &-& \left(1- 2 \, \gam \, \frac{G_N \, M}{r c^2} + ...\right)\,r^2 \,(dr^2
     \sin^2 \theta \, d\p^2+d\theta^2) \; .
\lb{eq:expansion} \eeqa

\in The coefficients $\alp$, $\beta$, $\gam$ are called the post
Newtonian parameters and their values depend on the considered
theory of gravity: in GR, one has $\alp=\beta=\gam=1$. The post
Newtonian parameters can be measured not only in the Solar System
but also in relativistic binary neutron stars such as
$PSR~1913+16$.\\ \in A generalization of the previous formalism is
known as {\sl Parametrized Post-Newtonian} (PPN) formalism.
Comparing metric theories of gravity among them and with
experimental results becomes particularly easy if the PPN
approximation is used. The following requirements are needed:

\begin{itemize}
 \item particles are moving slowly with respect to the speed of light;
 \item gravitational field is weak and considered as a perturbation of the flat spacetime;
 \item gravitational field is also static, \ie it does not change with time.
\end{itemize}

\in The PPN limit of metric theories of gravity is characterized
by a set of 10 real-valued parameters; each metric theory of
gravity corresponds to particular values of PPN parameters. The
PPN framework has been used first for the analysis of Solar System
gravitational experiments, then for the definition and the
analysis of new tests of gravitation theory and finally for the
analysis
and the classification of alternative metric theories of gravity.\\
\in By the middle 1970's, the Solar System was no more considered
as the unique testing ground of gravitation theories. Many
alternative theories of gravity agree with GR in the
Post-Newtonian limit and thus with Solar System experiments;
nevertheless, they do not agree for other predictions (such as
cosmology, neutron stars, black holes and gravitational radiation)
for which the post-Newtonian limit is not adequate.  In addition,
the possibility that experimental probes, such as gravitational
radiation detectors,
 would be available in the future to perform extra-Solar System tests led to the abandon of the
 Solar System as the only arena to test gravitation theories.\\
\in The study of the binary pulsar $PSR~1913+16$, discovered by R.
Hulse and J. Taylor \cite{HT}, showed that this system combines
large post-Newtonian gravitational effects, highly relativistic
gravitational fields (associated with the pulsar) with the
evidence of an emission of gravitational radiation (by the binary
system itself). Relativistic gravitational effects allowed one to
do accurate measurements of astrophysical parameters of the
system, such as the mass of the two neutron stars. The measurement
of the orbital period rate of change agreed with the prediction of
the gravitational waves (GW) emission in the GR framework, in
contradiction with the predictions of most alternative theories,
even those with PPN limits identical to GR. However, the evidence
was not conclusive to rule out other theories since several
shortcomings remain, up to now, unexplained.

\subsection{Basics of General Relativity} \lb{sec:GR}

GR is currently the most successful relativistic theory of
gravitation, introduced by A. Einstein \cite{AEGR} in 1915 to
include in a natural way the EP and SR theory (1905). Einstein's
derivation of GR has been driven by theoretical criteria of
elegance and simplicity, not involving neither experimental nor
observational results. Nevertheless, he had soon to compare his
theory with experimental tests. Three important ones confirmed GR:
$i)$ the anomalous perihelion of Mercury (1915) \cite{will}; $ii)$
the deflection of light by the Sun (1919) \cite{eddington}; $iii)$
the gravitational red-shift of light -- measured by Pound-Rebka
experiment (1960) \cite{rebka}. GR is a metric theory where the
metric tensor is the only gravitational field
\cite{weinberg,wald,landau,MTW}.

  The Einstein tensor $\Gd$ is defined as

\beq
 \Gd \;=\; R\dmunu - \half \gd R
\lb{eq:tens_Eins} \eeq

\noin with $R\dmunu$ being the Ricci tensor computed by
contracting the Riemann tensor $R\dmunu = R^\sig_{\,\,\mu\sig\nu}$
and $R=R^\mu_\mu=\gu R\dmunu$ the curvature scalar. The Einstein
tensor is symmetric thanks to the features of  Ricci  and  metric
tensors. It verifies the Bianchi identity $\nabla^\mu \Gd=0$. The
second term in the r.h.s. of Eq.~(\ref{eq:tens_Eins})  is
divergence-free as the stress-energy tensor $T\dmunu$ because of
the conservation of energy-momentum.\\ \in Einstein equations
present, in a formal way, the fundamental relation characterizing
the spacetime description of  GR: {\it matter curves spacetime
around it}. \\ \in Such a relation connects two 4-dimensional
symmetric tensors of rank 2, through a proportionality relation:
the Einstein tensor $\Gd$, defined above as a contraction of the
Riemann tensor $R\dabgd$, describing the curvature of a generic
manifold, and the stress-energy tensor $T\dmunu$ which contains
densities and energy-momentum fluxes. This tensor depends on all
possible types of energy (kinetic, potential, pressure, etc...)
and  also on gravitational field energy. Therefore, Einstein
equations are non-linear since gravity terms appear on both sides
(source and effect).\\ \in Einstein equations can be obtained in
two different ways. The first one is the natural covariant
generalization of Poisson equation for the Newtonian gravitational
potential, \ie
 \beq
 \Gd \;=\; \chi \; T\dmunu \qquad \text{with} \qquad \chi=\frac{8 \, \pi \, G_N}{c^4} \; .
\lb{eq:Einstein} \eeq

\in The proportionality constant $\chi$ is computed by assuming
the convergence to the Newtonian theory of gravity in the limit of
weak, time--independent gravitational fields and slow motions.\\
\in Einstein equations are non-linear second-order differential
equations; there are ten independent equations  since both sides
are  two-index symmetric tensors  (the solution consists in
computing the ten unknown functions of the metric components);
nevertheless, the Bianchi identities give four additional
constraints on $R\dmunu$, so there are only six independent
relations in Eq.~(\ref{eq:Einstein}). It is not possible to solve
these equations in  general, but even particular solutions, like
the ones in vacuum,
 are difficult to compute without some simplifying assumptions.\\
\in The second way to derive Einstein equations comes from the
variation of the action  principle where the curvature scalar
invariant $R$ is considered. The GR action  is the
Hilbert-Einstein action \cite{landau}

\beq
 {\s} \;=\; -\frac{1}{16\pi G_N}\vol \, \left[R \, + {\l}_{m} \right]
\lb{eq:action} \eeq

\noin linear in the Ricci curvature scalar $R$, minimally coupled,
being the gravitational coupling the constant $G_N$, where $g$ is
the metric determinant and ${\l}_{m}$ is the perfect-fluid matter
Lagrangian density.

Assuming spherical symmetry -- Eq.~(\ref{eq:isotropic}) -- there
are standard coordinates such that

\beq
 d s^2 \;=\; e^{\nu} \, dt^2 - e^{\lam} \, dr^2 -r^2 \, (\sin^2 \theta \, d\p^2+ d\theta^2)
\lb{eq:spherical_symmetry} \eeq

\noin where $\nu$ and $\lam$ are two functions of $t$ and $r$ only.\\
\in The solution of the Einstein equations in vacuum is

\beq
 e^{\nu} \;=\; e^{-\lam} \, f(t) \qquad \text{and} \qquad e^{\lam}\;=\; \frac{1+a}{r}
\lb{eq:sol_spherical} \eeq

\noin where $f(t)$ is an arbitrary positive function of $t$ and $a$ is a constant.
The metric is time--independent: this is the Birkhoff's theorem. For instance, it
states that the explosion of a supernova does not produce any time dependent modification of
spacetime if it is a spherically symmetric explosion, \ie there is no GW production
\footnote{This is a significant limitation for the GW production as large asymmetries are rather rare.}.\\
\in Let us assume that the source of  gravitational field is a
mass located at the origin $r=0$. In order to recover the
Newtonian limit far away from the origin, one must assume $a/r= 2
\p = -2\,(G_N\,M)/ (r\,c^2)$. The corresponding solution is called
the ``Schwarzschild solution'' (in standard coordinates)

\beq
 d s^2 \;=\; B(r)\, dt^2 - \frac{dr^2}{B(r)} - r^2 \, (\sin^2 \theta \, d\p^2+d\theta^2) \qquad \text{with} \qquad B(r)
 = 1 - \frac{2 \, G_N \, M}{r \, c^2}
\lb{eq:solution_Schwarzschild} \eeq

\noin where $M$ is the mass of the central body.\\ \in Generally,
the Schwarzschild radius of the central mass, $R_S=
2\,G_N\,M/c^2$, is much smaller than the physical radius $R_0$ of
the body itself. Therefore the solution
(\ref{eq:solution_Schwarzschild}) is only valid in a region where
$r \ge R_0 > R_S$. However, one can also consider a compact object
with $R_0 < R_S$; in that case the surface $r=R_S$ separates the
spacetime into two regions. Such a compact object is called a
Schwarzschild black hole. In the exterior region, where $r > R_S$,
the coordinates $t$ and $r$ are respectively time--like and
space--like ($g_{tt}=B>0$ and $g_{rr}=-1/B <0$). When a light
source approaches the surface $r=R_S$, the Einstein gravitational
shift becomes so huge that the frequency of the light becomes zero
everywhere; the photons loose their energy and cannot be seen
anymore. The surface $r=R_S$ is the last surface that can be seen
from outside and it is thus called the ``horizon''. In the
interior region, where $r<R_S$ one has $g_{tt}=B<0$ and
$g_{rr}=-1/B >0$; it means that $t$ is a space--like coordinate
while $r$ is a time--like coordinate.

\subsubsection{The weak field approximation and the linearized Einstein
equations}

Replacing the Newtonian limit by a less restrictive hypothesis
leads to the weak field approximation: practically, the field is
still weak, but it is allowed to change in time and there is no
more restriction on the test particles motion. New phenomena are
associated with this hypothesis like the emission of gravitational
radiation  and the deflection of light. This framework allows one
to split the metric $\gd$ into two parts: the flat Minkowski
metric $\eta\dmunu = \text{diag}(c,-1,-1,-1)$ plus a perturbative
term $\hd$, assumed to be small. This linearized version of GR
describes the propagation of a symmetric tensor $\hd$ on a flat
background spacetime. So, the metric reads

\beq
 \gd \;=\; \eta\dmunu + \hd \qquad \text{with} \qquad \left|\hd\right| \ll 1 \, .
\lb{eq:metr_decomp} \eeq

\in As $h$ is small, one can neglect terms higher than the first
order in the perturbation $\hd$; in particular, one can
raise/lower indexes with $\eta\dmunu$ and $\eta\umunu$ as the
corrections are of higher order in the perturbation

\beq
 \gu \;=\; \eta\umunu - \hu \qquad \text{with} \qquad h\umunu \;=\; \eta^{\mu\rho}\eta^{\nu\sig} h_{\rho\sig} \; .
\eeq

\in The aim is to find the equations of motion to which the
perturbations $\hd$ obey by investigating the Einstein equations
to the first order. Inserting the new metric
(\ref{eq:metr_decomp}) in the Einstein tensor, we obtain \beq
 \Gd \;=\; \half \, \left( \pa_{\sig} \pa_{\nu} h^{\sig}_{\mu} + \pa_{\sig} \pa_{\mu} h^{\sig}_{\nu} -
 \pa_{\mu} \pa_{\nu} h - \hd- \eta\dmunu \, \pa_{\rho} \pa_{\sig} h^{\rho\sig} + \eta\dmunu h \right)
 \lb{eq:new_tensor_Eins}
\eeq

\noin where $h = \eta\umunu \hd = h^\mu_\mu$ is the trace of the
perturbation and $ =  \half \pa_{tt} - \pa_{xx} - \pa_{yy} -
\pa_{zz}$ is the d'Alembertian of the
flat spacetime, using from now on (unless  otherwise stated) geometrical units for which $c=1$.\\
\in The stress-energy tensor is computed at the 0-order in $\hd$:
the energy and the momentum have to be small too, according to the
weak field approximation and the lowest non-vanishing order in
$T\dmunu$ is of the same order of magnitude as the perturbation.
Therefore, the conservation law becomes $\pa^\mu T\dmunu = 0$.

\subsubsection{Gravitational Waves} \lb{GW}

 GW are weak ripples in the curvature of spacetime, produced by the
 motions of matter.
 They propagate at the speed of light. The linearized
 Einstein equations allow wave solutions, in a way similar to electromagnetism.
 These GW are transverse to the propagation direction and show two independent polarization states.\\
\in The new metric (\ref{eq:metr_decomp}) does not fix the
spacetime frame completely; two possible gauges can be applied in
addition to simplify the Einstein equations. Using the Lorentz
gauge $\pa_\mu h^\mu_\lam -\half \pa_\lam h=0$, the Einstein
equations (\ref{eq:Einstein}) are linearized and can be written as

\beq
 \Box \, \hd - \half \, \eta\dmunu \, \Box \, h \;=\; \frac{16\pi G_N}{c^4} \, T\dmunu \; .
 \lb{eq:general_GW}
\eeq

\in The trace-reversed perturbation is defined as

\beq
 \bar{h}\dmunu \;=\; \hd - \half \, \eta\dmunu \, h \; .
 \lb{eq:trace_renv}
\eeq

\in One can choose a frame in which the harmonic gauge condition,
$\pa_\mu \bar{h}^\mu_\nu = 0,$ is verified. Then, the Einstein
field equations (\ref{eq:Einstein}) become

\beq
 \Box \bar{h}\dmunu \;=\;  \frac{16\,\pi \,G_N}{c^4} \, T\dmunu \; .
 \lb{eq:GW_eq}
\eeq

\noin and, in vacuum, one has simply

\beq
  \Box \bar{h}\dmunu \;=\; 0 \; .
 \lb{eq:vacuum}
\eeq

\in These equations are similar to the electromagnetism field
equations and one can use the same method to solve them. Indeed,
looking at the Einstein equations in vacuum --
Eq.~(\ref{eq:vacuum}) --, one can note that they are in the form
of a wave equation for $\bar{h}\dmunu$ -- the D'Alembertian
reduces to the form $\Box =  \half \pa_{tt} - \nabla^2$.
Therefore, in the absence of matter, one looks for plane waves
solutions

\beq
 \bar{h}\dmunu \;=\; C\dmunu e^{ik_\sig x^{\sig}}
 \lb{eq:vacuum_sol}
\eeq

\noin where $C\dmunu$ is a constant and symmetric tensor of rank 2 and $k$
is a time-like vector, the wave vector. The plane wave in Eq.~(\ref{eq:vacuum_sol})
is a solution of the linearized equations in vacuum -- Eq.~(\ref{eq:vacuum}) -- if the
wave vector $k$ is null, \ie satisfies the condition $k^\sig k_\sig = 0$ and shows that
GW propagate to the speed of light.\\
\in The four conditions of the harmonic gauge $k_\mu C\umunu = 0$
lead to six independent components for the symmetric tensor
$C\dmunu$. As there are still some unused degrees of freedom, one
can make another gauge choice on the tensor $C\dmunu$:

\beqa
 C^\mu_\mu &=& 0 \quad \text{Traceless}  \, ;\\
 C^{0\mu}  &=& 0 \quad \text{Transverse} \, .
\eeqa

\in One has, in this way, the so called Transverse-Traceless (TT)
gauge. These five relations give four new constraints on
$C\umunu$ in addition to the harmonic gauge condition; therefore,
only two independent components remain in $C\umunu$. As the wave
is traceless, one can check
from Eq.~(\ref{eq:trace_renv}) that $\bar{h}\dmunu^{TT} = \hd^{TT}$.\\
\in Therefore, the general form of the symmetric tensor $C\dmunu$
is finally

\beq
 C\dmunu =\left(
 \begin{array}{cccc}
  0 & 0 & 0 & 0\\
  0 & C_{11} & C_{12} & 0\\
  0 & C_{12} & - \, C_{11} & 0\\
  0 & 0 & 0 & 0\\
 \end{array}
 \right)
\eeq

\in Let us define $C_{11} = h_+$ and $C_{12} = h_\times$; GW
appears to have two polarized states -- ``$+$'' and ``$\times$''
-- which modify the spacetime curvature in  different ways. In
tensorial form, one can write

\beq
 h \;=\; \left[ h_+ \left( \vec{e_1} \otimes \vec{e_1} - \vec{e_2} \otimes \vec{e_2} \right)
 + 2 \, h_\times \left( \vec{e_1} \otimes \vec{e_2} \right) \right] \, e^{i\omega( t - x / c )} \; .
\eeq

\in  Being $\vec{\xi}=(\xi^1,\xi^2,\xi^3)$  the separation between
two free particles and taking into account the geodesic deviation
\cite{wald} which describes the evolution of two free-falling
particles, if the GW propagates in the direction $x^3$,  only
$\xi^1$ and $\xi^2$ are involved in its passage.
 Assuming a  polarized GW, the integration of
the geodesic deviation equation gives:

\begin{itemize}
 \item Polarization ``+'' ($h_\times = 0$)~:
  \beq\left(
 \begin{array}{c}
  \xi^1 ( t )\\
  \xi^2 ( t )
 \end{array}
 \right)
  =\left(
 \begin{array}{cc}
  1 + \half \, h_+e^{i k^\sig x_\sig} & 0 \\
  0 & 1 - \half \, h_+e^{i k^\sig x_\sig}
 \end{array}
 \right)\left(
 \begin{array}{c}
  \xi^1 ( 0 )\\
  \xi^2 ( 0 )
 \end{array}
 \right)
  \eeq
 \item Polarization ``$\times$'' ($h_+ = 0$)~:
  \beq
  \left(
 \begin{array}{c}
  \xi^1 ( t )\\
  \xi^2 ( t )
 \end{array}
 \right)
  =\left(
\begin{array}{cc}
 1 & \half \, h_\times e^{i k^\sig x_\sig} \\
\half \, h_\times e^{i k^\sig x_\sig} & 1
\end{array}\right)
\left(\begin{array}{cc}
\xi^1 ( 0 )\\
\xi^2 ( 0 )
\end{array}\right)
 \eeq
\end{itemize}

\in  Let us consider now a test--mass ring (a massless and
free--falling set of particles) interacting with GW, lying in an
plane orthogonal to the direction of the wave propagation. Its
oscillations depend on the GW polarization.

\in After having found a solution to Einstein equations in vacuum,
let us solve now Eq.~(\ref{eq:GW_eq}) with a non-zero source term.
The solution is computed using the retarded Green function

\beq
 \bar{h}\dmunu \left( t, \vec{x} \right) \; = \; \frac{4 \, G_N}{c^4} \int_{\vec{y} \,
 \epsilon \, \text{Source}} \frac{1}{\left|\, \vec{x} \, - \, \vec{y} \, \right|} \, T\dmunu \,
 \left( t_r, \vec{y} \right) d^3\vec{y}
\label{eq:pot_retard} \eeq

\noin with $\left| \, \vec{x}-\vec{y} \, \right| = \sqrt{\delta_{ij}(x^i-y^i)(x^j-y^j)}$
(Euclidean distance) and $t_r = t - \left| \, \vec{x}-\vec{y} \, \right| / c$ (retarded time).\\
\in Let us consider an isolated source with a density $\rho$ and a
characteristic dimension $\delta R$, located at a distance $R$ from
 the observation point $\vec{x}$. One assumes $\delta R \ll R$ so, in particular,
 one gets $\left| \, \vec{x}-\vec{y} \, \right| \approx R$ and one can move this constant term outside
 the integral in Eq.~(\ref{eq:pot_retard}). As the stress-energy tensor
 verifies the conservation of energy $\pa_\mu T\umunu = 0$, the harmonic
 gauge condition $\pa_\mu \bar{h}^\mu_\nu = 0$ is also verified. Moreover,
 the radiation is mostly emitted at frequencies $\omega / 2\pi$, so that
 ${\displaystyle \frac{\delta R}{ c} \ll \frac{1}{\omega}}$. Then, it is
 possible to demonstrate that only the spatial coordinates
 of  tensor $\bar{h}\dmunu$
 are different from zero.\\
\in The quadrupole momentum tensor $q_{ij}$ of the source energy
density is defined as

\beq
 q_{ij}(t) \;=\; \int_{\vec{y} \, \epsilon \,
 \text{source}} \, y_i \, y_j \, T_{00}( t, \vec{y} ) \, d^3\vec{y} \qquad
 \text{with } \qquad T_{00} \;\approx\; \rho c^2 \; .
\eeq

\in The metric perturbation is given by

\beq
 \bar{h}_{ij}(t,{\bf x}) = \frac{2 \, G_N}{R \, c^4} \, \ddot{q}_{ij} (t_r) \; .
\lb{eq:h_bar} \eeq

\in So, GW, generated by an isolated non-relativistic object, are
proportional to
 the second derivative of the quadrupolar momenta of the energy density. Eq.~(\ref{eq:h_bar})
 shows that the metric perturbation amplitude $h$ varies as
 the inverse of  distance to the source $R$; a faster decreasing with the distance, \eg $1/R^2$,
 would make vain the hope of any GW detection. Fortunately,
 GW detectors are sensitive to the amplitude of the signal.\\
\in The energy emitted by gravitational radiation is difficult to
define. A way to overcome this difficulty is to define the
stress--energy tensor by developing the metric $\gd$ and the
Einstein tensor $\Gd$ at the second order:

\beqa
 \gd &=& \eta\dmunu + \hd + h^{(2)}\dmunu         \; ,             \\
 \Gd &=& G^{(1)}\dmunu \; [\eta + h^{(2)}] + G^{(2)}\dmunu \; [\eta + h] \; .
\eeqa

\in Einstein equations in the vacuum $\Gd = 0$ can be written in
the form

\beq
 G^{(1)}\dmunu \; [\eta + h^{(2)}] \; = \; \frac{8 \, \pi \, G_N}{c^4} \; t\dmunu
\eeq

\noin with the definition

\beq
 t\dmunu \; = \;  \frac{c^4}{8 \, \pi \, G_N} \; G^{(2)}\dmunu \; [\eta + h] \; .
\eeq

\in The Bianchi identity says that $\pa_\mu t\umunu = 0$,
therefore $t\dmunu$ can be considered as the stress-energy tensor
of a gravitational field -- yet, it is only a pseudo-tensor
\cite{landau}. One can compute the energy density $t_{00}$ by
averaging over many cycles (because the energy is not localized)
the GW energy:

\beq
 t_{00} \; = \; \frac{c^2}{16 \, \pi \, G} \; \langle \; \dot{h_+}^2 + \dot{h_\times}^2 \; \rangle \; .
\eeq

\in Then, the source luminosity is ${\displaystyle L = r^2
\int_{\text{Sphere of radius r}} c \, t_{00} \, d\Omega}$.
Introducing the reduced quadrupolar momenta

\beq
 Q_{ij} \; = \; q_{ij} - \frac{1}{3} \delta_{ij} \delta^{kl} q_{kl}
\eeq

\noin one obtains the  Einstein quadrupole formula

\beq
 L \; = \; \frac{G}{c^5} \; \langle \; \frac{d^3 Q_{ij}}{dt^3} \frac{d^3 Q_{ij}}{dt^3} \; \rangle \; .
\lb{eq:quadrupole} \eeq

\in It is worth noticing that a GW emission requires a variation
of the quadrupolar momentum, as shown already in
Eq.~(\ref{eq:h_bar}). This typical feature of  gravitational
radiation, added to the weakness of the coupling constant between
gravitation and matter -- here
$G/c^5~\approx~10^{-53}~\text{W}^{-1}$ -- explains why GW
amplitudes are so small compared to those produced by
electromagnetic radiation. In conclusion, the gravitational
radiation is quadrupolar and a symmetric spherical body does not
emit GW because its reduced quadrupolar momenta are zero.\\ \in
The corresponding quantum field is the graviton with zero mass and
spin $2$.
 Within the framework of more general theories, the gravitation can be described
 as a combination of states with a spin $2$ and spin zero or as a
 particle with spin maximum $2$. One can also imagine that the mass
 of graviton is not zero or that the state mass spectrum is complex.
 Presently, there is no observational reason to doubt that
 the present observational bounds on the mass of the graviton are
 severe.\\
\in Discovering GW would open {\sl ``a new window onto the
Universe''}  \cite{thorne}. It is clear that being sensitive to an
additional radiation would lead to major discoveries like when the
Universe became observed through radio, X or gamma waves. Then, it
would allow physicists to test GR in the strong field regime, to
check the gravity velocity (assumed to be the speed of light in
the Einstein theory) or to verify that GW only change distances
perpendicular to their direction of propagation.\\ \in Alternative
relativistic theories of gravity also predict the existence of GW.
However, many essential features of the radiation are different:
the number of polarization states, the propagation speed, or the
efficiency of wave generation. We shall not take into account GW
any further in this review, but it is worth stressing that their
detection is of fundamental interest  since their features and
parameters, through the PPN limit or in strong field regime, allow
a discrimination among relativistic theories of gravity
\cite{thorne}.

\subsection{Discussion of Classical Tests of General Relativity}
\lb{sec:discussion}

Historically, the gravitational red--shift experiment, the
deflection of light and the anomalous perihelion shift of Mercury
are considered as the three classical tests of GR \cite{will}:
indeed, they were among the first observable effects of GR to be
computed by Einstein. The gravitational red-shift experiment can
be considered as a test of the EEP, upon which GR and other metric
theories of gravity are founded. Moreover, a further test has been
introduced: the time delay of light due to the presence of
gravitational field.\\

The deflection of light by the Sun is described, in physical
units,  by the deflection angle

\beq
 \delta \theta \;=\; \left(\frac{1+\gam}{2} \right) \frac{4\,M}{D} \, \left(\frac{1+ \cos \,
 \theta_0}{2}\right)
\lb{eq:deflection_variation} \eeq

\noin where $\theta_0$ is the angle between the unperturbed paths
of the photons from the source, the Sun, and the reference source
and $D$ is the distance between the center of  mass and the point
of closest approach of the unperturbed ray. For $\gam=1$ one finds
the results of GR.\\ \in The explanation of the anomalous
perihelion shift of Mercury orbit was the subject of a controversy
to establish if this shift was a confirmation or a refutation of
GR because of the apparent existence of a Solar quadrupole
momentum that could contribute a fraction of the observed
perihelion shift.\\ \in The time delay of light depends on the
strength created by the gravitational field of a massive body
located near the photon path: because of this effect, a light
signal will take a longer time to cross a given distance than it
would if the Newtonian theory was valid. The time delay is

\beq
 \delta t \;=\; 2 \, (1+\gam) \, M \, \ln \left(\frac{4 R_{\oplus} R_p}{R_{\odot}^2}\right)
\lb{eq:delay} \eeq

\noin where $M$ is the mass of the body, $R_{\oplus}$ the Earth
radius, $R_{\odot}$ the Solar radius and $R_p$ the distance of a
planet or  a spacecraft.\\ \in Beside experimental tests, it is
worth mentioning some theories about tests: kinematic frameworks
\cite{robertson,MS,VS,tourrenc} to prove the Local Lorentz
Invariance (LLI), dynamical frameworks \cite{LL,blanchet} to test
WEP, LLI and
 Local Position Invariance (LPI), unification theory \cite{DP}
 motivated to test WEP, LLI and LPI.

 \newpage
 Tab.I shows the PPN parameters and their significance in GR, in
semi--conservative theories and in scalar--tensor theories
\cite{will}.
\begin{table}[h!]
\begin{center}
\bet{|c|c|c|c|c|} \hline
          &Meaning &Value&Value in         & Value in\\
Parameter &relative to GR  &in GR&semi-conservative& scalar-tensor \\
          &                &     &theories         & theories\\ \hline
 $\gam$   &The quantity of space-curvature&$1$&$\gam$&${\displaystyle \frac{1+\omega}{2+\omega}}$ \\
          &produced by unit rest mass&  &    & \\ \hline
 $\beta$  &The quantity of ``non-linearity''&$1$&$\beta$&$1+\lam$ \\
          & in gravity            &   &       & \\ \hline
  $\xi$   &existence of &$0$&$\xi$&$0$ \\
          &preferred-location effects           &   &        & \\\hline
 $\alp_1$ &existence of                         &$0$&$\alp_1$&$0$\\
 $\alp_2$ &preferred-frame effects              &$0$&$\alp_2$&$0$\\
 $\alp_3$ &                                     &$0$&$0$&$0$ \\ \hline
 $\zeta_1$&violation of                         &$0$&$0$&$0$\\
 $\zeta_2$&total momentum                       &$0$&$0$&$0$\\
 $\zeta_3$& conservation                        &$0$&$0$&$0$\\
 $\zeta_4$&                                     &$0$&$0$&$0$\\ \hline
\eet
\end{center}
\end{table}
A direct test of the WEP is the $E\ddot{o}tv\ddot{o}s$ experiment,
\ie the comparison of the rest acceleration of two
laboratory-sized bodies of different composition in an external
gravitational field. Indeed, the accelerations of different bodies
are not equal if the WEP is not valid. A measurement or limit on
the relative difference in acceleration yields to a quantity
called ``$E\ddot{o}tv\ddot{o}s$ ratio'' given by

\beq
 \eta \; \equiv \; 2 \, \frac{|a_1-a_2|}{|a_1+a_2|}
\lb{eq:eta_parameter} \eeq

\noin where $a_1$ and $a_2$ are the accelerations  of the two
bodies.\\ \in Experimental limits on $\eta$ place limits on the
WEP-violation parameter. Several high-precision
$E\ddot{o}tv\ddot{o}s$-type experiments have been performed and
limits on $\eta$ from different experiments are shown in Tab.II
\cite{will}.
\begin{table}[h!]
\begin{center}
\bet{|c|c|c|} \hline
 Experiment             & Method              & Limit on $|\eta|$  \\
 \hline
 Newton                 & pendula             &$10^{-3}$           \\
 Bessel                 & pendula             &$2 \times 10^{-5}$  \\
 E$\ddot{o}$tv$\ddot{o}$s & torsion balance     &$5 \times 10^{-9}$\\
 Potter                 & pendula             &$2 \times 10^{-5}$  \\
 Renner                 & torsion balance     &$2 \times 10^{-9}$  \\
 Princeton              & torsion balance     &$10^{-11}$          \\
 Moscow                 & torsion balance     &$10^{-12}$          \\
 Munich                 & free fall           &$3 \times 10^{-4}$  \\
 Stanford               & magnetic suspension &$10^{-4}$           \\
 Boulder                & flotation on water  &$4 \times 10^{-11}$ \\
 Orbital                & free fall in orbit  &$10^{-15}\div10^{-18}$ not yet performed
 \\\hline \eet
\end{center}
\end{table}
Very few experiments have been used to make quantitative tests of
the LLI in the same way that E$\ddot{o}$tv$\ddot{o}$s experiments
have been used to test the WEP. There is one experiment that can
be interpreted as a ``clean" test of the LLI: the Hughes-Drever
experiment, performed independently in $1959 \div 60$. If the LLI
is violated, then there must be a preferred rest frame, presumably
the one associated to the Cosmic Microwave Background (CMB), in
which the local laws of physics take their special form.
Deviations from this form would depend on the velocity of the
laboratory relative to the preferred frame. New experimental tests
will be performed in the future, using atomic clocks, macroscopic
resonators or laser interferometry, for instance the LATOR mission
\cite{lator} which aims to carry out a test of the curvature of
the Solar System's gravity field with an accuracy of $10^{-8}$.
The two principal tests of LPI are gravitational red-shift
experiments testing the existence of spatial dependence on the
outcomes of local experiments, and the measurements of the
possible time variations of the fundamental non--gravitational
constants. Some Solar System experiments testing the SEP are based
on the Lunar-Laser-Ranging (LLR) E$\ddot{o}$tv$\ddot{o}$s
experiment which takes into account a potential Nordtvedt effect
(not yet observed), \ie the violation of the GWEP. The breaking of
the WEP for massive self-gravitating bodies (GWEP), which several
metric theories predict, has a variety of observable consequences:
the most important one  is a polarization of the Moon orbit about
the Earth. Because the Moon self-gravitational
 energy is smaller than  the Earth one, the Nordtvedt effect causes
 that  Earth and Moon
 fall toward the Sun, with slightly different acceleration. The ``Nordtvedt'' parameter
 $\eta$ and its limits are shown in Tab.III Its expression
 is $\eta = 4 \beta -\gam -3-(10/3)\xi -\alp_1+
 (2/3)\alp_2-(2/3)\zeta_1-(1/3)\zeta_2$. In GR, $\eta=0$.\\
\in Tests to establish preferred-frame and preferred-location
effects can be divided into two main categories: geophysical tests
and orbital tests.
\begin{itemize}
 \item Geophysical tests: some metric theories of gravity predict preferred-location
 effects considering  the locally estimated gravitational constant $G_L$, measured using
 Cavendish--like experiments. These effects are violations of SEP. Limits are shown in Tab.III.
 \item Orbital tests: there are a number of observable effects of preferred-frame
 and preferred-location types in the orbital motions of bodies
 governed by the $n$-body equations of motion. The most important
  ones are the perihelion shifts of planets coming in addition to the ``classical'' shift.
\end{itemize}

\newpage
\in The current PPN parameter limits are shown in Tab.III.
\begin{table}[h!]
\begin{center}
\bet{|c|c|c|c|} \hline
 PPN Parameter &Effect          & Limit & Source \\ \hline
 $\gam -1$     &time delay    & $2.3(5.7) \times 10^{-5}$&Cassini Doppler  \\
               &light deflection& $3 \times 10^{-4}$ &VLBI \\
               &                & Unpublished data & \\
               &                & (Eubanks \text{\etal} \cite{will}) & \\
 $\beta-1$       &perihelion shift&$3 \times 10^{-3}$&$J_2=10^{-3}$ \\
               &Nordtvedt effect&$5 \times 10^{-4}$&LLR, $\eta = 4 \beta -\gam -3$\\
   $\xi$         &Earth tides     &$\times 10^{-3}$&gravimeters \\
   $\alp_1$      &orbital         &$\times 10^{-4}$&LLR\\
               &polarization&$2 \times 10^{-4}$&LLR\\
   $\alp_2$      &spin precession &$4 \times 10^{-7}$& sun axis\\
   $\alp_3$      &self acceleration&$2 \times 10^{-20}$&$\dot{P}$ statistics\\
   $\eta$        &Nordtvedt effect&$3 \times 10^{-4}$&LLR \\
   $\zeta_1$     &     $-$          &$2 \times 10^{-2}$&combined PPN bounds \\
   $\zeta_2$     &binary acceleration&$4 \times 10^{-5}$&$\ddot{P}$ for PSR 1913+16 \\
   $\zeta_3$     &Newton's 3rd law &$10^{-8}$&Lunar acceleration \\
   $\zeta_4$     &      $-$       &$-$& not independent \\\hline\eet
\end{center}
\end{table}
\in Fig.1 shows some experimental tests performed during a period
of eighty years about the parameter ${\displaystyle
\left(\frac{1+\gam}{2}\right)}$. It is possible to derive a bound
on $\omega>20,000$ in scalar-tensor gravity \cite{will}.
\begin{figure*}[ht]
\centering \resizebox{10cm}{!}{\includegraphics{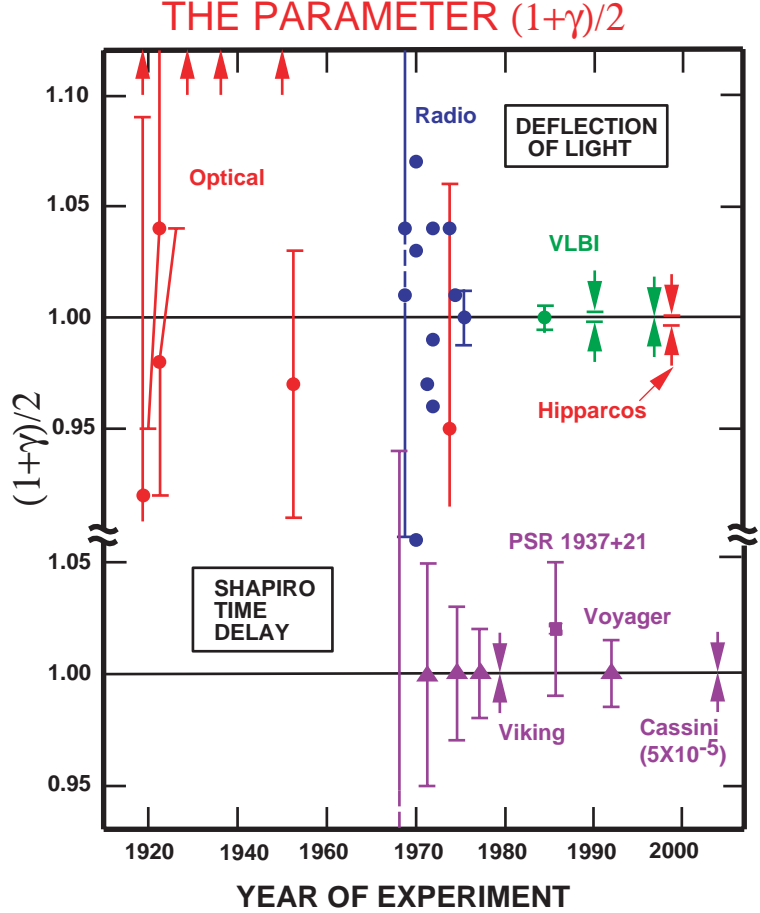}}
\hfill \caption{Measurements of the coefficient $(1+\gam)/2$ from
light deflection and time
 delay measurements -- the GR value is unity. Shapiro time-delay measurements using Viking
  spacecraft and Very Large Baseline Interferometry (VLBI) light deflection
  measurements are in agreement with GR to $0.1 \%$. Results from Cassini
  mission are in agreement at the level of $5 \times 10^{-5}$.}
\end{figure*}
From a quick view of limits and data in the tables, it is clear
that no definite answer can be given on which is the "{\it true}"
theory of gravity. GR works very well in weak field limit but
shortcomings arise as soon as quantum fields, cosmology or strong
gravitational fields are considered. Moreover, no definite tests
have been performed outside Solar System and the "{\it fact}" that
GR works at larger scales (\eg galactic sizes) is nothing else but
a pure assumption (issues as "missing" or "dark" matter could be
read as "indications" that GR does not work and further
ingredients are needed to explain dynamics). Due to these reasons,
alternative theories of gravity have to be studied and, possibly,
tested.

\section{Extended theories of gravity}
\lb{alternative}
 Extended Theories of Gravity (ETG) are, by definition, all the approaches to the physics
 of gravitational interaction which enlarge the Einstein scheme
 through an addition of corrective terms into the Hilbert-Einstein
 action (\ref{eq:action}). There are several reasons to do so
and the resulting field equations are modified  leading to different physical predictions.\\
\in From a conceptual point of view, there is {\it a priori} no
reason to restrict the gravitational Lagrangian to a linear
function of the Ricci scalar $R$, minimally coupled with matter
\cite{francaviglia}. Furthermore, the idea that there are no
``exact'' laws of physics but that the Lagrangians of physical
interactions are ``stochastic'' functions -- with the property
that local gauge invariances (\ie conservation laws) are well
approximated in the low energy limit and that physical constants
can vary -- has been taken into serious consideration -- see
Ref.~\cite{ottewill} for instance. This scheme was adopted in
order to deal with the quantization on curved spacetimes and the
result was that the interactions among quantum scalar fields and
background geometry or the gravitational self--interactions yield
corrective terms in the Einstein--Hilbert Lagrangian
\cite{birrell}. Moreover, it has been realized that such
corrective terms are inescapable if we want to obtain the
effective action of quantum gravity on scales closed to the Planck
length \cite{vilkovisky}. Higher--order terms in curvature
invariants (such as $R^{2}$, $R\umunu R\dmunu$,
$R^{\mu\nu\alp\beta}R_{\mu\nu\alp\beta}$, $R \,\Box R$, or $R
\,\Box^{k}R$) or non-minimally coupled terms between scalar fields
and geometry (such as $\p^{2}R$) have to be added in the effective
Lagrangian of gravitational field when quantum corrections are
considered. For instance, one can note that such terms occur in
the effective Lagrangian of strings or in Kaluza--Klein theories,
when the mechanism of dimensional reduction is used \cite{veneziano}.\\
\in Using a completely different point of view, these extended
theories become interesting when one tries to incorporate the Mach
principle in gravity: this principle states
  that the local inertial frame is determined by some average of the motion of distant astronomical
  objects \cite{bondi}, so that gravitational coupling can be
  scale-dependent and related to some scalar field. As a consequence,  the concept
  of ``inertia'' and equivalence principle have to be revised. For example, the
  Brans--Dicke theory is a serious attempt to define an alternative theory to the Einstein gravity:
  it takes into account a variable Newton gravitational constant,
  whose dynamics is governed by a scalar field
  non-minimally coupled with geometry. In such a way, the Mach principle is better
  implemented \cite{brans,cimento,sciama}.\\
\in Beside fundamental physics motivations, all these theories
have acquired a huge interest in cosmology due to the fact that
they "naturally" exhibit inflationary behaviours able to overcome
the shortcomings of Standard Cosmological Model (based on GR). The
related cosmological models seem very realistic and, several
times, capable of matching with the observations
\cite{starobinsky,la}.\\
 \in Furthermore, it is possible to show
that, via conformal transformations, the higher--order and
non-minimally coupled terms always correspond to the Einstein
gravity plus one or more than one minimally coupled scalar fields
\cite{teyssandier,maeda,wands,conf,gottloeber}. More precisely,
higher--order terms always
 appear as a contribution of order two in the equations of motion.
 For example, a term like $R^{2}$ gives fourth
 order equations \cite{ruzmaikin}, $R \ \Box R$ gives sixth order equations \cite{gottloeber,sixth},
 $R \,\Box^{2}R$ gives eighth order equations \cite{eight} and so on. By a conformal transformation,
 any 2nd--order term corresponds to a scalar field: for example, fourth--order gravity gives Einstein
 plus one scalar field, sixth order gravity gives Einstein plus two scalar
 fields and so on \cite{gottloeber,schmidt1}.\\
This feature results very interesting if we want to obtain
multiple inflationary events since an early stage could select
``very'' large-scale structures (clusters of galaxies today),
while a late stage could select ``small'' large-scale structures
(galaxies today) \cite{sixth}. The philosophy is that each
inflationary era is connected with the dynamics of a scalar field.
Furthermore, these extended schemes naturally could solve the
problem of "graceful exit" bypassing the shortcomings of former
inflationary models \cite{la,aclo}.\\
 \in In this section, we want to discuss some of these theories,
outlining their main features. In particular, we shall consider:
\begin{itemize}
 \item GR with a cosmological constant, the minimal extension discussed by Einstein himself;
 \item higher-order-scalar-tensor theories, which are the general
 scheme embracing all the ETG;
 \item scalar-tensor theories of gravity, discussing, in particular,
 the most famous example which is the Brans-Dicke theory;
 \item fourth-order theory of gravity;
 \item higher than fourth--order theories of gravity.
\end{itemize}
As a general feature (apart the first case), we shall see that the
Einstein field equations result modified in two senses: $i)$
geometry can be nonminimally coupled to some scalar field, and/or
$ii)$ higher than second order derivative terms in the metric come
out.

\subsection{General Relativity with a cosmological constant}
\lb{sec:cosmo_constant}

The determination of the cosmological constant \cite{lambdat} is
one of the most challenging issue of modern physics since by
fixing its value, one could contribute to select self--consistent
models of fundamental physics and cosmology. Briefly, its
measurement should provide the gravity vacuum state
\cite{weinberg1}, should allow to understand the mechanism which
led to the today observed large scale structure from the early
universe \cite{guth,linde}, and to predict what will be the fate
of the whole universe (``no--hair conjecture'' -- see below)
\cite{hoyle}.\\ \in The presence of a cosmological constant $\Lam$
modifies the Einstein equations (\ref{eq:Einstein}); this term has
been first included by Einstein himself and tuned to obtain a
static Universe model. Later he rejected this ``fine tuning'' as
his ``biggest blunder'' after the discovery by Hubble of the
expansion of the Universe by looking at red-shifted galaxy
spectra. Ironically, recent measurements \cite{snIa,WMAP}  seem to
favor a non--zero value of $\Lam$ which could be the major
contribution to the cosmological density.\\ \in Assuming the
existence of a cosmological constant $\Lam$, Einstein equations
(\ref{eq:Einstein}) can be written as

\beq
 \Gd  = \frac{8 \, \pi \, G}{c^4} \, T\dmunu +\Lam \, \gd
\lb{eq:Einstein_general} \eeq

\noin and the action in presence of a cosmological constant is
simply

\beq
 {\s} = -\frac{1}{16\pi G_N}\vol \left[R \, + 2 \Lam \right] \; .
\lb{eq:cosmo_action} \eeq

\in On the right-hand side of the Einstein equations, this
$\Lam$-term  is a stress-energy tensor contribution automatically
conserved by definition of the metric connection. It can be
considered as a {\sl vacuum energy density}, a source of energy
and momentum present also in  absence of matter fields; this
interpretation agrees with quantum field theory which predicts
that the vacuum should have some sort of energy and momentum.
However, this prediction is $50$ orders of magnitude above the
current measurements of $\Lam$!\\ \in Thanks to the recent
measurements of distant supernovae Ia luminosity \cite{snIa}, a
non-zero cosmological constant $\Lam$ is coming back in the
footlights. Nevertheless, other possibilities are still
investigated to explain the origin of this ``dark energy'', like
the {\sl quintessence}. Let $\Omega_\Lam$ be the energy density
associated to the cosmological constant, normalized to the
critical density of the Universe. Taking into account the recent
results of  WMAP experiment \cite{WMAP}, its value is

\beq
 \Omega_\Lam \;simeq\; 0.73 \pm 0.04
\lb{eq:Omega_Lam} \eeq

\noin with the hypothesis $\Omega_{\text{total}} = 1$, well confirmed by experimental data.\\
\in From the Eq.~(\ref{eq:Omega_Lam}) it is possible to deduce a
value for $\Lam$

\beq
 \Lam \;=\; (1.29 \pm 0.15) \, 10^{-52}
 \left( \frac{H_0}{ 71 \, \text{km} \text{s}^{-1} / Mpc} \right)^2 \; \text{m}^{-2}
\lb{eq:force} \eeq

\noin where $H_0$ is the Hubble constant predicted by WMAP
experiment  \cite{WMAP}. The ratio  between the gravitational
force per mass unit due to a mass $M$, located at a distance $D$,
and therefore due to the cosmological constant is given by
\cite{kenyon}

\beq
 \Upsilon \; = \;  \frac{G_N M}{c^2 \Lam D^3} \; = \; 4
 \times 10^{21} \left( \frac{10^{-52}}{\Lam} \right)
 \left( \frac{\text{Earth-Sun distance}}{D} \right)^3 \left( \frac{M}{M_\odot} \right) \gg 1
\lb{} \eeq

\noin as the total gravitational force per unit mass  is
${\displaystyle F = - \frac{ G_N M }{ D^2 } + \frac{ c^2 \Lam D }{
3 }}$. The cosmological constant $\Lam$ is not negligible for
cosmological distances.\\ \in The no--hair conjecture \cite{hoyle}
claims that if there is a positive cosmological constant,  models
of expanding universe will approach a {\sl de Sitter} behavior,
\ie if a cosmological constant is present, the universe will
necessarily become homogeneous and isotropic whatever its initial
conditions were. However, there is no general proof of such a
conjecture and there are even theoretical counter--examples of
initially expanding and then re-collapsing universes which never
become de Sitter \cite{cotsakis}.\\ \in On the other hand, a
simplified version of the conjecture can be proved \cite{waldPRD}:

\begin{quote}
 {\sl All Bianchi  cosmologies (except Bianchi IX), in presence of a positive
 cosmological constant,
 asymptotically approach the de Sitter behaviour}.
\end{quote}

\in It is worthwhile noticing that the cosmological constant is a
true constant (added by hands) and that the contracted Bianchi
identity  is not used in the proof, which is thus independent of
the matter evolution. In order to extend the ``no--hair
conjecture'' to extended theories of gravitation, one has to
introduce different sets of conditions (with respect to those
given in \cite{waldPRD}) since the cosmological constant is not
given, {\sl a priori}, but can be recovered from the dynamics of
scalar fields (considering higher--order geometric terms in the
gravitational Lagrangian also as a kind of scalar fields
\cite{wands,schmidt1}).\\ \in Such conditions must not use the
energy conditions \cite{hawking}, but they have to allow the
introduction of a kind of {\sl effective cosmological constant}
$\Lam_{eff}$, which asymptotically becomes the de Sitter constant
\cite{lambdat,lambdastep}. This feature is due to the fact that,
in an expanding universe, all the contributions to the energy
density and to the Ricci tensor must decay as some power of the
scale factor $a(t)$, while the cosmological constant is the only
term which does not decrease with time. Hence, in an expanding
universe, $\Lam$ is the asymptotically dominant term in the
Einstein equations giving rise to a de Sitter spacetime. Actually,
the effective cosmological constant is time--dependent but, at the
end, it has to coincide with the de Sitter one (\ie the true
gravitational vacuum state). Then, given any extended theory of
gravity, it could be possible, in general, to define an effective
time-varying cosmological constant which becomes the ``true''
cosmological constant if and only if the model asymptotically
approaches de Sitter (\ie if the no--hair conjecture is recovered
asymptotically). This fact introduces constraints on the choice of
gravitational coupling, of  scalar field potential and of
higher--order geometrical terms whose combinations are used as
components of the effective stress--energy tensor. Terms like
$R^2$, $\Box R$ and $\p$ can be all treated as ``scalar fields" in
the construction of $\Lam_{eff}$, \ie all of them give rise to
extra--terms in the field equations which contribute to the
construction of an effective stress--energy tensor
$T\dmunu^{(eff)}$.

\in In this context, a fundamental issue is to select the classes
of gravitational theories and the conditions which "naturally"
allow to recover an effective time--dependent cosmological
constant without considering special initial data.\\ In the
framework of extended gravity theories, the meaning of a time
dependent ``cosmological constant" can be discussed
\cite{lambdat}. As we said, the main request is that such
effective cosmological constants evolve (at least asymptotically)
toward the actual cosmological constant which means that the de
Sitter behaviour has to be recovered.

\subsection{Higher-order-scalar-tensor theories: the general scheme}
\lb{sec:4dim}

If the gravitational Lagrangian is nonlinear in the Ricci scalar
(and, in general, in the curvature invariants) its dynamics,
(\ie the Einstein equations), has an order higher than second; for
this reason, such theories are often called {\it higher--order gravitational theories}.
 Physically, as we said, they are interesting since higher--order terms in curvature invariants
 appear when one performs a one--loop renormalization of matter and gravitational fields
 in a curved background (see for example Ref.~\cite{birrell,barth}).\\
\in We can take into account the the most general class of
higher--order-scalar--tensor theories in four dimensions. They can
be generated by the action

\beq
 {\s} \;=\; \vol \, \left[F(R,\Box R,\Box^{2}R,..\Box^kR,\p) -\frac{\epsilon}{2} \, \gu
 \p \ddemu \p \ddenu+{\l}_{m}\right]
\lb{eq:high_action} \eeq

\noin where $F$ is an unspecified function of curvature invariants
and of a scalar field $\p$. The term ${\l}_{m}$ is, as above, the
minimally coupled ordinary matter contribution. Physical units,
$8\pi G=c=\hbar=1$,  will be used, unless otherwise stated;
 $\epsilon$ is a constant which specifies the kinetic term of the theory.\\
\in The field equations are obtained by varying
Eq.~(\ref{eq:high_action}) with respect to the metric $\gd$. One
gets

\beqa
 \Gu &=&\frac{1}{{\g}}\left[T\umunu +\half \, \gu \, (F-{\g}R)+
   (g^{\mu \lam}g^{\nu\sig}-\gu g^{\lam \sig}) \, {\g}_{;\lam \sig}+\right.\no\\
    & & +\half \sum_{i=1}^{k}\sum_{j=1}^{i}(\gu g^{\lam \sig}+
   g^{\mu\lam} g^{\nu\sig})(\Box^{j-i})_{;\sig}
   \left(\Box^{i-j}\frac{\pa F}{\pa \Box^{i}R}\right)_{;\lam}+\no\\
    & &\left.-\gu g^{\lam\sig}\left((\Box^{j-1}R)_{;\sig}
   \Box^{i-j}\frac{\pa F}{\pa \Box^{i}R}\right)_{;\lam}\right]
\lb{eq:high_field_eq} \eeqa

\noin with

\beq
 {\g}\equiv\sum_{j=0}^{n}\Box^{j}
 \left(\frac{\pa F}{\pa \Box^{j} R}\right)
\lb{eq:high_field_values} \eeq

\noin $\Gu$, as above, is the Einstein tensor so, considering the
r.h.s. of Eqs.(\ref{eq:high_field_eq},) all the extra
contributions can be dealt as stress-energy tensor terms.
Eqs.~(\ref{eq:high_field_eq}) are differential equations of order
$(2k+4)$. The stress--energy tensor $T\umunu$  contains the
kinetic part of the scalar field and of the ordinary matter

\beq
 T\dmunu \;=\; T^{(m)}\dmunu+\frac{\epsilon}{2} \, [\p\ddemu\p\ddenu-\half \, \p\udea\p\ddea]\; .
\lb{eq:high_stress} \eeq

\in The (eventual) contribution of a potential $\v$ is contained in the definition of $F$.
 From now on, a capital $F$ will indicate a Lagrangian density containing also the contribution of
 a potential $\v$ while $f(\p)$, $f(R)$, or $f(R,\Box R)$ will indicate functions
 of such fields without  potential terms.\\
\in By varying the action ${\s}$ with respect to the scalar field
$\p$, one obtains the Klein--Gordon equation

\beq \epsilon \, \Box\p \;=\; -\frac{\pa F}{\pa\p} \; .
\lb{eq:high_KG_eq} \eeq

Dynamics is completely assigned taking into account
Eqs.(\ref{eq:high_field_eq}), (\ref{eq:high_KG_eq}) and Bianchi
identity for $T\dmunu$.

\in Several approaches can be used to treat such equations. For
example, by a conformal transformation, it is possible to reduce
an extended theory of gravity to a (multi) scalar--tensor theory
of gravity \cite{wands,gottloeber,schmidt1,damour}. The theories
studied below are specific examples coming from the general scheme
which we have outlined here.

\subsection{Scalar--tensor gravity} \lb{sec:scalar--tensor}

The {\sl scalar--tensor theories of gravity} are  theories, in
which both the metric tensor $\gd$ and a fundamental scalar field
$\phi$ are involved \cite{cimento}. Their action can be recovered
from the above (\ref{eq:high_action}) with the choice \beq
 \lb{eq:scalar-tensor_general}
 F(\p, R) \;=\; f(\p) \, R-\v \qquad \text{and} \qquad \epsilon=-1
\lb{eq:scalar-tensor_positions} \eeq where the coupling $f(\p)$
and the potential $\v$ are generic functions of the scalar field
$\phi$. The increasing interest in such theories is mainly due to
the success of the inflationary paradigm in solving the Standard
Cosmological Model shortcomings. Moreover, the presence of a
scalar field is relevant also in
 multidimensional gravity, especially in Kaluza-Klein theories and in
 the effective action of String Theory.\\
\in In this framework, the strength of gravity, given by the local
value of the gravitational coupling, is different from place to
place and time to time. For instance, the Brans-Dicke theory (the
most famous scalar-tensor theory \cite{brans}),  includes the
hypothesis suggested by Dirac of the variation of the
gravitational coupling with time \cite{sciama}.\\ \in All these
scalar--tensor theories  do not satisfy the SEP as the variation
of the gravitational constant $G_{eff}$ -- which is, in general,
different from $G_N$, the standard Newton gravitational constant
-- implies that local gravitational physics depends on the scalar
field. Theories which have such a feature are called {\sl
non-minimally coupled theories}.\\ \in Some Lagrangians
corresponding to particular scalar-tensor theories of gravity are:

\begin{itemize}
 \item the effective string-dilaton Lagrangian \cite{veneziano}

   \beq
    {\l} \;=\; \sqrt{-g} \, e^{-\p}\left(R +  g\umunu \p \ddemu \p \ddenu - \Lam \right)
   \lb{eq:string_lagrangian}
   \eeq
   which we shall discuss in details below;

\item the Callan-Coleman-Jackiw Lagrangian
\cite{CCJ} in which ${\displaystyle f(\p)=\half-\frac{\p^2}{12}}$;
 \item the conformally coupled theory \cite{penrose}

   \beq
    {\l} \;=\; \sqrt{-g} \left( -\frac{\p^2}{6} R + \half \, g\umunu \p \ddemu \p\ddenu - \v \right)
   \lb{eq:coupled_lagrangian}
   \eeq

   \noin where $\v$ is generic and ${\displaystyle
   f(\p)=-\frac{\p^2}{6}}$.
\end{itemize}
These Lagrangians are are all particular cases of
(\ref{eq:scalar-tensor_general}).

\in In these theories,  the gravitational coupling is determined
by the form of $f(\p)$. We can have two physically interesting
situations which could be tested by experiments:
\begin{itemize}
 \item[1)] $G_{eff}(\p(t))_{t \ra \infty} \ra G_N$, the standard Newton gravitational
 constant is asymptotically recovered and GR restored;
 \item[2)] the possibility that gravitational coupling is not asymptotically constant, \ie $G_{eff}$ is
 always
 varying with the epoch  and $\dot{G}_{eff}/G_{eff}\vert_{now}\neq 0$.
\end{itemize}

\in The variability of $G$ can be actually tested by three classes
of experiments:

\begin{itemize}
 \item the information coming from the Sun dynamics.
 \item The observations based on LLR
 combined with the information obtained studying the Sun.  As we saw above,
 LLR consists in measuring the round-trip
 travel time and thus the distance between a transmitter and a reflector located on the Moon;
 the change of round-trip time contains information on the Earth-Moon system.
 The round-trip travel time has been investigated for many years; combining these
 data with those coming from the Sun evolution and the Earth-Mars radar ranging,
 the more update estimation of $\dot{G}/G$ ranges from $0.4 \times 10^{-11}$ to $10^{-11}$
 per year \cite{dickey,uzan}.
 \item The binary pulsar systems.
 In order to extract data from his type of system, it has been necessary
  to extend the post-Newtonian approximation, which can be applied only
  in the presence of a weakly gravitationally interacting n-body system,
  to strong gravitationally interacting systems. The estimation of $\dot{G}/G$ is $2 \times 10^{-11}$ per
  year \cite{uzan,damouresp}.
\end{itemize}
Moreover, gravitational lensing observations are going to become a
further class of tests for the variability of $G_N$
\cite{krauss}\\
\in There is another fundamental issue, coming from the basis of
metric theories, to take into account scalar-tensor gravity.
According to the Mach principle, the local inertial frame is
determined by the average of the motions of distant astronomical
objects \cite{bondi}. Trying to incorporate this principle into
metric theories of gravity, Brans and Dicke \cite{brans} have
constructed a theory of gravity where  the total matter
distribution influences the strength (and the coupling) of
gravitational field in each point of spacetime. Therefore, a
scalar field $\p$ is introduced as a new gravitational variable
together with the standard metric tensor $\gd$. The Brans-Dicke
theory is the first and simplest example of a scalar-tensor theory
of gravity described  by the Lagrangian density

\beq
 {\l} \;=\; \sqrt{-g} \, \left(\p \, R - \omega \, \frac{\p \ddemu \p \udenu}{\p} + {\l}_{m}\right)
 \lb{eq:scalar-tensor_Lagrangian}
\eeq

\noin where $\omega$ is a dimensionless parameter and ${\l}_{m}$,
as usual, is the Lagrangian density of  matter including all the
non-gravitational fields.\\ \in Let us now   perform a conformal
transformation $\gd \ra {\bar{g}\dmunu} = \lam \, \gd$ on the
Lagrangian (\ref{eq:scalar-tensor_Lagrangian}). The field $\lam$
is proportional to the scalar field $\p$, \ie $\lam=G_0\p$ where
$G_0$ an arbitrary constant. The Lagrangian
(\ref{eq:scalar-tensor_Lagrangian}) is transformed as

\beq
 \bar{{\l}} \;=\; \sqrt{-\bar{g}} \, \left(\bar{R} + G_0 \, \bar{{\l}}_{m} + G_0 \, \bar{{\l}}_{\lam} \right)
 \lb{eq:scalar-tensor_Lagr}
\eeq

\noin where

\beq
 \bar{{\l}}_{\lam} \;=\; -\left(\frac{3+2 \, \omega}{4 \, \pi \, G_0 \, \lam^2}\right) \, \lam\udemu \lam\ddenu
\lb{eq:lagr_density} \eeq

\noin and $\bar{{\l}}_{m}$ is the conformally transformed
Lagrangian density of  matter. Einstein equations can be written
as

\beq
 \bar{R}\dmunu - \half \, \bar{g}\dmunu \, \bar{R} \;=\; G_0 \, \bar{\tau}\dmunu
 \lb{eq:Ein_eqs}
\eeq

\noin where the stress-energy tensor is the sum of two
contributions

\beq
 \bar{\tau}\dmunu \;=\; T^{(m)}\dmunu + \Lam\dmunu \, (\lam^{field})
 \,,
 \lb{eq:Ein_tensor}
\eeq

so that all the contributions due to the scalar field $\lambda$
(or $\phi$) result as a further term in the stress-energy tensor.

\in This approach constitutes  an alternative to GR and  several
agreements with experimental results demonstrated that GR is not
the unique possible theory of gravity.\\ \in Moreover, the
transformed form of  scalar-tensor theory of gravity has the
advantage to be more manageable, being similar to Einstein
standard description. Nevertheless, this form presents some
drawbacks, \eg the rest mass is not constant in the conformal
transformed frame, so the photons trajectories are still geodesics
but the massive particles equations of motion are modified.
 This
new approach to gravitation has increased the interest in
scalar-tensor theories of gravity and, especially, in those in
which the gravitational constant is variable. The Newtonian limit
of induced--gravity theories, where the scalar field is
non-minimally coupled to the geometry, strictly depends on the
parameters of  coupling and  self-interaction potential as it has
to be by a straightforward PPN parametrization \cite{will}.

Considering the choice (\ref{eq:scalar-tensor_general}), the most
general action of a theory of gravity where a scalar--field is
non--minimally coupled to the geometry is, in four dimensions
\cite{cqg}, \beq
 {\s} \;=\; \vol \, \left[f(\p) R +\half \, \gu \p \ddemu \p \ddenu-\v+{\l}_{m}\right] \; .
\lb{eq:high_scalar-tensor_action} \eeq

\in It is possible to obtain the field equations by varying the
action (\ref{eq:high_scalar-tensor_action}) with respect to the
metric tensor field $\gd$. Therefore, Einstein equations are

\beq
 \Gd \;=\; \tilde{T}\dmunu \;=\; -\frac{1}{2 \, f(\p)} \, T\dmunu \;=\; T^{(eff)}\dmunu \; .
\lb{eq:high_scalar-tensor_field_values} \eeq

 The stress--energy tensor is defined as

\beq
 T\dmunu \;=\; T^{(\p)}\dmunu+T^{(m)}\dmunu \;=\; -2 \, f(\p) \, T^{(eff)}\dmunu
\lb{eq:stress_scalar-tensor} \eeq

\noin \ie $T\dmunu$ includes the scalar--field  and the standard
matter terms; in Eq.~(\ref{eq:high_scalar-tensor_field_values})
and Eq.~(\ref{eq:stress_scalar-tensor}), $T^{(eff)}\dmunu$ is the
{\it effective stress-energy tensor} which contains terms of
non-minimal coupling, kinetic terms, the potential of  scalar
field $\p$ and the usual stress-energy tensor of matter,
$T^{(m)}\dmunu$, calculated varying ${\l}_{m}$ with respect to the
metric.  It is

\beqa
  T^{(eff)}\dmunu \;=\; \frac{1}{f(\p)} \, \left\{ -\half \,
        \p\ddemu \p\ddenu  + \frac{1}{4}\, \gd g\uab \p\ddea \p\ddeb \right.
         - \half \, \gd \, \v -\gd \Box f(\p) + \no \\ + \left.
        f(\p) \ddemunu -4\,\pi \,\tilde{G} \, T^{(m)}\dmunu \right\} \;,
        \lb{eq:T_eff_expression} \eeqa
        $\tilde {G}$ is a dimensional,
strictly positive, constant. \in Terms coming from the coupling
$f(\p)$, which were outside $T\dmunu$ in
Eq.~(\ref{eq:high_field_eq}), have been assembled. Here ${\g}$,
defined in Eq.~(\ref{eq:high_field_values}), is given by ${\g}
=f(\p)$. The standard Newton gravitational constant is replaced by
the effective coupling

\beq
 G_{eff}\;=\;-\frac{1}{2 f(\p)}
\lb{eq:newton_constant} \eeq

\noin which is, in general, different from $G_N$; $f(\p)$ is the
generic function describing the scalar--field coupling. In units
with $c=1$, Einstein gravity is obtained when the scalar field
coupling $f(\p)$ is a constant which value is $-\half$ and
$\tilde{G}$ reduces then to the Newton gravitational constant
$G_{N}$ \cite{cimento}.\\ \in The corresponding Klein--Gordon
equation is

\beq
 \Box \, \p - R f'(\p) +\vp \;=\; 0 \; .
\lb{eq:scalar-tensor_KG_eq} \eeq

\in The derivation of such an equation from the contracted Bianchi
identity for $T\dmunu$  is discussed in \cite{cimento,nmc,cqg}.
 As a general feature, the models
described by Eq.~(\ref{eq:high_scalar-tensor_action}) can be
singularity free \cite{quartic}; then, there are no restrictions
on the interval of time on which the scale factor $a(t)$ and the
scalar
field $\p(t)$ can be defined.\\
\in Field equations derived from the action
(\ref{eq:high_scalar-tensor_action}) can be recast in a
Brans--Dicke equivalent form by choosing

\beq
 \pv \; = \; f(\p) \qquad \omega\left(\pv\right)=\frac{f(\p)}{2\,f'(\p)^2}
 \qquad W \left(\pv\right)=V\left(\p(\pv)\right) \; .
\lb{eq:choice} \eeq

\in The action (\ref{eq:high_scalar-tensor_action}) now reads

\beq
 {\s} \; = \; \vol \, \left[\pv \, R  +\frac{\omega(\pv)}{\pv}\, g\umunu \pv\ddemu \pv  \ddenu-W(\pv)+{\l}_{m}\right]
 \lb{eq:B-D_action}
\eeq

\noin which is nothing else but the extension of the original
Brans--Dicke proposal (where $\omega$ is a constant), with
$\omega=\omega(\pv)$ plus a potential term $W(\pv)$. The sign of
$\omega$ is chosen according to those of coupling and potential.
By varying the new action with respect to $\gd$ and the new scalar
field $\pv$, one obtains again the field equations. The effective
stress-energy tensor in Eq.~(\ref{eq:T_eff_expression}), in the
same units as before, becomes

\beqa
 T^{(eff)}\dmunu \;=\; -\frac{4\,\pi \,\tilde{G}}{\pv}\,T^{(m)}\dmunu -
 \frac{\omega(\pv)}{\pv^2} \left(\pv \ddemu \pv \ddenu - \half\, \gd g\uab \pv\ddea \pv \ddeb \right)+
 \no \\ + \frac{1}{\pv} \left(\pv\ddemunu -\gd \Box \pv\right)-\frac{1}{2\,\pv}\, \gd  W\left(\pv\right)
 \lb{eq:T_eff_new}
\eeqa

\noin and the Klein--Gordon equation
(\ref{eq:scalar-tensor_KG_eq}) becomes

\beq
 2\,\omega \left( \pv\right) \Box\pv- \frac{\omega\left(\pv\right)}{\pv}\,
 g\uab \pv\ddea \pv \ddeb  -\frac{d\omega\left(\pv\right)}{d\pv}\, g\uab \pv\ddea
 \pv \ddeb -\pv R +\pv \, W'\left( \pv \right) \;=\; 0 \; .
\lb{eq:KGeqs} \eeq

\in This last equation is usually rewritten eliminating the scalar
curvature term $R$ with the help of Eq.~(\ref{eq:T_eff_new}), so
that one obtains

\beq
  \Box\pv \; = \; \frac{1}{3-2\,\omega(\pv)} \left(-4\,\pi \, \tilde{G} \, T^{(m)}- 2\,
  W\left(\pv\right)+ \pv\,W'\left(\pv\right)- \frac{d  \omega \left(\pv\right)}{d \pv}\,
  g\uab \pv\ddea \pv \ddeb \right) \; .
\lb{eq:KGeqs_eqs} \eeq

\in The minus sign in the denominator comes from the sign
convention chosen in the action (\ref{eq:B-D_action}). The
linearized equations can be derived from the action
(\ref{eq:high_scalar-tensor_action}) or, equivalently, from
(\ref{eq:B-D_action}) as we shall see in the next section.

\subsection{Higher-order gravity} \lb{sec:fourth-order}

Higher--order theories can be reduced to minimally coupled
scalar--tensor ones, and {\it vice--versa}, by a conformal
transformation \cite{maeda}. The simplest case to be considered is

\beq
 {\l} \;=\; \sqrt{-g} \, f(R)
\lb{eq:fouth_order_lagrangian} \eeq

\noin where $g$ is the determinant of  metric $\gd$ and $R$ is the
curvature scalar. This Lagrangian is recovered from the extended
action (\ref{eq:high_action}) with the choice $F=f(R)$ and
$\epsilon=0$, that is

\beq
 {\s} \;=\; \vol \, \left[f(R)+{\l}_{m}\right]
\lb{eq:fourth_order_action}\,. \eeq

 This action depends on a generic function of the Ricci scalar so
 the Hilbert-Einstein  GR action is just the linear case --
see Eq.~(\ref{eq:action}). Once more, geometric units
$8~\pi~G=c=\hbar=1$ are used. From the Lagrangian
(\ref{eq:fouth_order_lagrangian}) one gets the fourth-order
equations

\beq
 f'(R) \, R\dab-\half f(R)g\dab \;=\; f'(R)\udemunu\left(g_{\alp \mu}g_{\beta\nu}-g\dab\gd\right)+T^{(m)}\dmunu
\lb{eq:fourth_order}\,, \eeq

 \noin which are fourth-order equations, because of the term
$f'(R)\udemunu$. The prime indicates  the derivative with respect
to $R$ -- the standard Einstein  equations are immediately
recovered if $f(R)=R$.

The introduction of a new set of variables

\beq
 p\;=\; f'(R) \;=\; f'(\gd,\pa_\sig \, \gd,\pa_{\sig\rho \gd}) \qquad \text{and} \qquad \bar{g} \;=\; p \, \gd
\lb{eq:variables} \eeq

\noin is interpreted as the {\sl Einstein frame} ($p$, $\gd$
variables), where the word ``frame'' means ``change in variables''
and the word ``Einstein'' means that using the transformation $g
\ra ({p,\bar{g}})$ the non-standard equations
(\ref{eq:fourth_order}) are transformed into something closer to
the standard form, also called the Einstein form. Indeed, the
(${\bar{g}\dmunu,p}$) Einstein equations, without considering
matter terms, are

\beq
 \bar{G}\dmunu \;=\; \frac{1}{p^2} \, \left[ \frac{3}{2} \, p_{,\mu} p_{,\nu} -
 \frac{3}{4} \bar{g}\dmunu \bar{g}\uab p_{,\alp} p_{,\beta} + \half \, \bar{g}\dmunu (f(R) - R f'(R)) \right] \; .
\eeq

\in Defining

\beq
 \p \;=\; \sqrt{\frac{3}{2}} \, \ln p
\eeq

\noin one gets

\beq
 \bar{G}\dmunu \;=\; \left[\p_{,\mu} \p_{,\nu} - \half \, \bar{g} \dmunu \p_{,\sig} \p^{,\sig} + \bar{g} \dmunu \v \right]
\lb{eq:Einstein_fourth} \eeq

\noin where $\v$ is defined as

\beq
 \v \;=\; \half \, \frac{[f(R) - R \, f'(R)]}{f'^2 (R)} \Big|_{R=R(p(\p))}
\lb{eq:v_def} \eeq

\noin and where the function $R=R(p(\p))$ can be defined thanks to
the condition $f'(R) \neq 0$. Eq.~(\ref{eq:Einstein_fourth})
corresponds to the Lagrangian

\beq
 \bar{\l} \;=\; \sqrt{-\bar{g}} \left( - \half \, \bar{R} + \half \, \bar{g}\umunu \p_{,\mu} \p^{,\nu} - \v \right)
\eeq

\noin which describes the gravitational interaction with a  self-gravitating, "minimally" coupled,
 massive scalar field.\\

 Eq.~(\ref{eq:fourth_order}) can also be
written in the above Einstein form ${\displaystyle
\Gd=\tilde{T}\dmunu}$ by defining

\beq \tilde{T}\dmunu \;=\; \frac{1}{f'(R)} \,\left\{\half \,
\gd\left[f(R)-R \, f'(R)\right]+f'(R)_{;\mu\nu}-\gd\,\Box
f'(R)+T^{(m)}\dmunu \right\} \; . \lb{eq:fourth_order_stress} \eeq

\in In the framework of higher--order--scalar--tensor theories,
another choice is extensively studied  \cite{maeda,lambdat}

\beq
 F \;=\; F(R,\p) \qquad \mbox{any}\;\epsilon \qquad \text{and} \qquad {\l}_{m}=0
\lb{eq:fourth_order_scalar-tensor_positions} \eeq

\noin with the action

\beq
 {\s} \;=\; \vol \, \left[F(R,\p) -\frac{\epsilon}{2} \gu \p\ddemu \p\ddenu\right] \; .
\lb{eq:fouth_order_scalar-tensor_action} \eeq

\in The Einstein equations are

\beq
 \Gd \;=\; \frac{1}{\g} \, \left\{T\umunu+\half\,(F-{\g}R)+\left[{\g}\ddemunu-\gd\Box {\g}\right]\right\}
\lb{eq:fourth_order_scalar-tensor_eq} \eeq

\noin where ${\g}$ and $T\dmunu$ are

\beq
  {\g} \;\equiv\; \frac{\pa F}{\pa R} \qquad \text{and}
  \qquad T\dmunu \;=\; \frac{\epsilon}{2} \, [\p\ddemu\p\ddenu-\half\,\p\udea\p\ddea]\;.
\lb{eq:order_fourth} \eeq

\in The (eventual) contribution of a potential $\v$ is contained
in the definition of $F$. By varying ${\s}$ with respect to the
scalar field $\p$, one obtains the Klein--Gordon equation of the
form (\ref{eq:high_KG_eq}).

Finally, a pure higher than fourth--order gravity theory is
recovered with the choice

\beq
 F \;=\; F(R,\Box R) \qquad \epsilon \;=\; 0 \qquad \text{and} \qquad {\l}_{m} \;=\;0
\lb{eq:high_fourth_positions} \eeq

\noin which is, in general, an eighth--order theory. If $F$
depends only linearly on $\Box R$, one has a sixth--order theory.
In this model, the action (\ref{eq:high_action}) becomes

\beq
 {\s} \;=\; \vol \, F(R,\Box R) \;.
\lb{eq:high_fourth_action} \eeq

\in The Einstein field equations are now

\begin{equation}
 \Gu \;=\; \frac{1}{2 \g} \, \left\{ \gu[F-{\g}R]+ 2{\g}\udemunu- 2\gu \,
 \Box {\g}- \gu \, [{\fo}_{;\gam}R^{;\gam}+{\fo}\Box R] + [{\fo}^{;\mu}R^{;\nu}+{\fo}^{;\nu}R^{;\mu}] \right\}
\lb{eq:high_fourth_einstein_tensor}
\end{equation}

\in where

\beq
 {\g} \;=\; \frac{\pa F}{\pa R}+\Box\frac{\pa F}{\pa \Box R} \qquad
 \text{and} \qquad {\fo}=\frac{\pa F}{\pa \Box R} \; .
\lb{eq:high_fourth_positions_values} \eeq

\in Another generalization of the Hilbert-Einstein action  is
obtained by including scalars of more than second order in
derivatives of the metric \cite{nmdc}. The action for these
theories has the form

\beq
 {\s} \;=\; \int d^nx \,\sqrt{-g} \, \left[f(\p) R + \alp_1 \,
 R^2 + \alp_2 \, R\umunu R\dmunu + \alp_3 \, \gu \, \nabla_\mu \, R \, \nabla_\nu \, R + ...\right]
 \lb{eq:higher_order}
\eeq

\noin where the $\alp$ coefficients are coupling constants and the
final suspension points represent every other scalar invariants
which one can assemble from the curvature tensor, its contractions
and its derivatives. This kind of actions are particularly
interesting in one-loop quantum gravity since well represent
self-gravity interactions and scalar field interactions
\cite{odintsov}.

\section{Newtonian limit in extended theories of gravity} \lb{newtonian-limit}

As discussed in Sec.II, several relativistic theories {\it do not}
match perfectly GR theoretical
 results but generalize them by introducing corrections to the Newtonian potential
  which could have interesting physical consequences. For example, some theories give
   rise to terms capable of explaining the flat rotation curve of galaxies without using
   dark matter -- for instance the fourth--order conformal theory
   proposed by Mannheim \etal~\cite{mannheim}. Others use Yukawa corrections to the Newton potential
   for the same purpose \cite{sanders}. Besides, indications of an apparent, anomalous, long--range
   acceleration revealed from the data analysis of Pioneer 10/11, Galileo, and Ulysses spacecraft
   trajectories could be framed in a general theoretical scheme by taking into account Yukawa--like or
   higher-order corrections to the Newtonian potential \cite{anderson}.\\
\in In general, any relativistic theory of gravitation can yield
corrections to the Newton potential which, in the PPN formalism,
could furnish tests for this theory \cite{will}. In this section,
the Newtonian limit is investigated for fourth-order and
scalar-tensor theories showing that several new features come out
with respect to the corresponding GR Newtonian limit.

\subsection{Recovering the Newtonian limit in fourth-order theories of gravity}
Particularly interesting, among the fourth-order theories of
gravity, are the so called $R^2$-theories \cite{ruzmaikin}, where
the inclusion of terms proportional to $R \dmunu R \umunu $ and
$R^2$ in the gravitational action gives rise to a class of
effectively multi-mass models of gravity \cite{stelle}; such terms
also produce a stabilization of the divergence structure of
gravity, allowing it to be renormalized through its matter
couplings. The empty space solutions of Einstein equations, $R
\umunu =0$, are also solutions of the field equations derived from
actions like $\int \sqrt{-g} \, R \dmunu R \umunu $ and $\int
\sqrt{-g} \, R^2$, but the classical tests of GR are not
automatically satisfied. In fact, although \eg the Schwarzschild
solution satisfies the empty space equations, it is not the one
that couples to a positive definitive matter distribution; those
solutions of purely fourth-derivative models which do couple to a
positive matter source are not asymptotically flat at infinity.\\
\in Let us consider only models derived from actions including
both the Hilbert
 action $\int \sqrt{-g}\, R$ and the four derivative terms.
 There are only two independent possible additions, because of the
 Gauss-Bonnet relation in four-dimensions, so there is only a
 two-parameter family of field equations without cosmological term (see also \cite{birrell}).\\
\in The higher-order action is, in this case,

\beq
 {\s} \;=\; - \int \, \sqrt{-g} \, (\alp \, R \dmunu R \umunu - \beta \, R^2 +\gam \, k^{-2} \, R)
\lb{eq:stelle_action} \eeq

\noin where $k^2=32 \, \pi \, G_N$, and $\alp$, $\beta$ and $\gam$
are dimensionless numbers. It  turns out that the correct physical
value for $\gam$ is 2, as in the Einstein theory, restoring the
standard coupling $(16\pi G_N)^{-1}$.\\ \in As we shall see, the
static spherically symmetric solutions to the field equations
derived from Eq.~(\ref{eq:stelle_action}) either reduce
asymptotically to the sum of a Newtonian and two Yukawa
potentials, or they are unbounded at infinity and must be
eliminated by some proper boundary conditions. The masses, in
these Yukawa potentials, are only weakly constrained by the
observational evidence to be $\geq 10^{-4}~cm^{-1}$ (in
geometrical units). Although the magnitude of these effects is
negligible at laboratory distances, there are some interesting
features which could result interesting at astronomical distances
beyond the Solar System. Coupling the linearized
 theory to a pressurized fluid distribution shows that the coefficients of the Yukawa potentials
 depend on the pressure and the size of the distribution. This shows that Birkhoff's
 theorem   is not valid in these models.\\
\in The Yukawa potentials and the failure of Birkhoff theorem in
the static case are complemented by the existence of "massive"
radiation terms in the dynamical field. The analysis of the field
dynamics, both in the action and in the interaction between
sources, shows that the gravitational field has eight degrees of
freedom, two corresponding to the massless spin-two graviton, five
to a massive spin-two excitation and the last one to a massive
scalar. This decomposition is carried out into various individual
spin components using the TT decomposition of a symmetric tensor.
 The massive excitations are separated from the massless ones
  by introducing auxiliary oscillator variables, in terms of
  which the Lagrangian may be separated into a sum of Lagrangians for massive scalar-fields.
  The massive spin-two field comes with a minus sign relative to the other fields,
  both in the oscillator variable Lagrangian and in the radiation.
  Classically, the corresponding excitation has negative energy;
  therefore, there is a breakdown of causality which can cause
  superpositions of waves with group velocities greater than the speed of light.\\
\in The field equations obtained by the action
(\ref{eq:stelle_action}), including matter, are

\beqa
 H\dmunu \;=\; \left(\alp-2 \beta\right)\, R_{;\mu ;\nu} - \alp \,
  R_{\mu \nu ;\eta}^{;\eta} - \left(\half \, \alp - 2 \, \beta \right) \,
  \gd \, R_{\mu \nu ;\eta}^{;\eta} + 2 \, \alp \, R^{\eta \lam} R_{\mu \eta \nu \lam} \lb{eq:stelle_field} \\
-2 \, \beta \, R \, R \dmunu - \half \, \gd \, \left(\alp \,
R^{\eta \lam} R_{\eta \lam} - \beta \, R^2\right) + \gam \, k^{-2}
\, R \dmunu - \half \, \gam \, k^{-2} \gd \, R \;=\;  \half \, T
\dmunu
 \lb{eq:stelle_field_1}
\eeqa

\in Eq.~(\ref{eq:stelle_field}) and Eq.~(\ref{eq:stelle_field_1})
are linked by generalized Bianchi identities  $H^{\mu \nu}_{ ;\nu}
\equiv 0$. In order to extract the physical content of
Eq.~(\ref{eq:stelle_field}) for the static spherically symmetric
gravitational field, one can work in Schwarzschild coordinates,
see Eq.~(\ref{eq:isotropic}).
\noin The above equations can be
linearized assuming $A(r)=1+V(r)$ and $B(r)=1+W(r)$ and keeping
only terms linear in $V$ and $W$.\\ \in The general solution to
the linearized field equations (\ref{eq:stelle_field}), \ie $H
\dmunu^L=0$, is

\beqa
 V &=& C + \frac{C^{2,0}}{r} + C^{2+} \,\frac{e^{m_2 \,r}}{r} + C^{2-}
  \,\frac{e^{-m_2 \,r}}{r} + C^{0+} \,\frac{e^{m_0 \,r}}{r} + C^{0-}
  \,\frac{e^{-m_0 \,r}}{r} \lb{eq:stelle_solV} \\
 W &=& - \frac{C^{2,0}}{r} - C^{2+} \,\frac{e^{m_2 \,r}}{r} - C^{2-}
 \,\frac{e^{-m_2 \,r}}{r} + C^{0+} \,\frac{e^{m_0 \, r}}{r} + C^{0-}
 \,\frac{e^{-m_0 \,r}}{r} +\no \\
 &+& \half C^{2+} m_2 \, e^{m_2 \,r} - \half C^{2-} m_2 \, e^{-m_2 \,r} -C^{0+} m_0
 \, e^{m_0 \,r} + C^{0-} m_0 \, \frac{e^{-m_0 \,r}}{r}
\lb{eq:stelle_solW} \eeqa

\noin where $V$ is the gravitational (Newton-like) potential and
$m_2$ and $m_0$ are related to the coupling constant
$\alpha,\,\beta,\,\gamma$ and $k$ as

\beq
 m_2\;=\; \sqrt{\frac{\gam}{\alp \, k^2}}\,, \qquad \text{and}
 \qquad m_0\;=\;\sqrt{\frac{\gam}{2\,(3\,\beta-\alp) \, k^2}}\,,
\eeq

\noin $C$, $C^{2,0}$, $C^{2+}$, $C^{2-}$, $C^{0+}$, $C^{0-}$ are
arbitrary constants. The rising exponentials $r^{-1} e^{mr}$ occur
in Eq.~(\ref{eq:stelle_solV}) and Eq.~(\ref{eq:stelle_solW})
because they are  solutions of the static Klein-Gordon equation
$(\nabla^2-m^2) \, \p=0$.  As in the Klein-Gordon case, they must
be eliminated by imposing boundary conditions at infinity.
Counting the number of parameters that must be determined by
source coupling and boundary conditions at infinity, the parameter
$C$ should be discarded,
 since it may be absorbed into a trivial rescaling of the time coordinate $t$. Consequently,
 the general solution to the linearized static spherically symmetric equations is a
 five-parameter family, before the imposition of boundary conditions.\\
\in In the case of a massive point particle with $T \dmunu =
\delta_\mu^0 \,\delta_\nu^0 \, M \,\delta^3(x)$, the gravitational
potential takes the form

\beq
 V = - \frac{k^2 \,M}{8 \, \pi \, \gam r} + \frac{k^2 \,M}{6 \,\pi \,\gam}
 \frac{e^{-m_2 \,r}}{r} - \frac{k^2 \,M}{24 \, \pi \, \gam} \frac{e^{-m_0 \,r}}{r} \; .
\lb{eq:stelle_point} \eeq

\in The comparison at infinity with the Newtonian result
$V=-2~G~M~r^{-1}$ shows that the correct physical value of $\gam$
is $2$. The higher-derivative terms can produce an appreciable
effect also at small distances from the source of the field, at a
scale set by $m_2$ and $m_0$. At the origin, the Newtonian term
$1/r$ is cancelled and
 Eq.~(\ref{eq:stelle_point}) tends to the finite value $k^2~M~(24~\pi~\gam)^{-1}(m_0 - 4~m_2)$.\\
\in If, instead of a point particle, one couples to an extended
source with some structure, the peculiarities of the static field
coupling in these models will be revealed. One can consider now a
spherical distribution of a perfect fluid with an internal
pressure that is maintained by an elastic membrane. The stress
tensor for this source is

\beq
 T \dmunu =\left(
 \begin{array}{cccc}
  P &              0                &             0                & 0 \\
  0  & P\,[1-\half \,l\,\delta\,(r-l)]\,r^2 &            0         & 0 \\
  0  & 0  & P\,[1- \half \, l \, \delta\,(r-l)]\,r^2\, \sin^2 \theta & 0 \\
  0  &          0     &       0           & 3\,M\,(4\,\pi\,l^3)^{-1} \\
 \end{array}\right)
\lb{eq:stelle_tensor} \eeq

\noin where $l$ is the radius of the source, $M$ is its total
energy and $P$ is the internal pressure. The delta functions in
$T_{\theta \theta}$ and $T_{\p \p}$ correspond to the membrane
tension and are necessary to ensure equilibrium of the source,
given by the Bianchi identities $T^{\mu \nu}_{;\nu}=0$, which
read, in spherical coordinates, as

\beq
 (T_{r r})'+ 2 \,r^{-1} \,T_{r r}-r\, T_{\theta \theta}-r\, \sin^2 \theta \, T_{\p \p}\;=\; 0 \; .
\lb{eq:T_spherical} \eeq

\in The exterior field, for the source
Eq.~(\ref{eq:stelle_tensor}), is

\beqa
 V_{fluid}^{ext} \;=\; - \frac{k^2 \, M}{8 \, \pi \, \gam r}
+\frac{k^2 e^{-m_2 \,r}}{\gam \, r} \, s_1 - \frac{k^2 e^{-m_0
\,r}}{2 \,\pi \,r} \, s_2 \lb{eq:fluid} \eeqa

\noin where $s_1$ and $s_2$ are defined as

\beqa
 s_1&=& \frac{M}{2 \,\pi \,l^3} \, \left[ \frac{l \, \cosh(m_2 \,l)}{m_2^2} - \frac{\sinh(m_2 \, l)}{m_2^3} \right]+ \no \\
 &-& P \, \left[ \frac{\sinh(m_2 \, l)}{m_2^3} - \frac{l \, \cosh(m_2 \,l)}{m_2^2} + \frac{l^2 \, \sinh(m_2 \, l)}{3 \, m_2} \right] \no \\
 s_2&=&\frac{M}{4 \,\pi \,l^3} \left[ \frac{l \cosh(m_0 \,l)}{m_0^2} - \frac{\sinh(m_0 \, l)}{m_0^3} \right] +\no \\
 &-&P \left[ \frac{\sinh(m_0 \, l)}{m_0^3} - \frac{l \, \cosh(m_0 \, l)}{m_0^2} + \frac{l^2 \, \sinh(m_0 \, l)}{3 \, m_0} \right] \; .
\eeqa

\in In Eq.~(\ref{eq:fluid}), the existence of the quantities $m_2$ and $m_0$ with the
 dimensions of mass has enabled the gravitational coupling to involve the source pressure and
 radius, as well as the total energy $M$. In the limit $l \ra 0$, the pressure dependence drops out
 and Eq.~(\ref{eq:fluid}) tends to the potential (\ref{eq:stelle_point}) for a point particle.
 The three-parameter family of static solutions -- Eq.~(\ref{eq:fluid}) --
 shows that the Birkhoff theorem
 does not hold for this models; this is the case  for the Yukawa potentials
 in Eq.~(\ref{eq:stelle_point}). Eq.~(\ref{eq:fluid}) corresponds to the virtual exchange of massive particles
 and has a real counterpart in massive spin-two and spin-zero radiation. Eq.~(\ref{eq:stelle_point})
  and Eq.~(\ref{eq:fluid}) give  an acceptable Newtonian limit only for real
   $m_2$ and $m_0$,
  so that there are only falling exponentials and not oscillating $1/r$ terms at infinity.
  This corresponds to the absence of tachyons (both positive and negative energy) in the dynamical field.
  It is also possible to remove the two massive fields by picking combinations of $\alp$ and $\beta$ that
   make $m_2$ or $m_0$ infinite, choosing, respectively, $\alp=0$ or $\alp=3\beta$ in Eq.~(\ref{eq:stelle_field}).\\
\in The static field (\ref{eq:stelle_point}) may be considered as
the gravitational potential of a star (or a generic spherical
body) and observations could be used to compute bounds on the
values of $m_2$ and $m_0$. However, astronomical tests in Solar
System  are useless here. In fact, by considering the motion of
Mercury, the orbital precession is known, up to now,  to about one
part in $10^9$. The corrections due to the higher-derivative terms
are of order $e^{-mr}$, where, for $r$, one can use the radius of
Mercury orbit $\sim 5 \times 10^6$ km. This gives a lower bound on
the masses of the order $\sim 5 \times 10^{-11}$ cm$^{-1}$. Since
there is no discontinuity in the coupling to light in the limit
$m_2$, $m_0 \ra \infty$, measurements of light bending by the Sun
do no better. Laboratory experiments on the validity of the
Newtonian $1/r^2$ force seem much more promising, as they can seek
agreement on the meter--scale. Yet, high-precision Cavendish-type
experiments have to be improved. On the other hand, such results
could provide interesting clues at extragalactic scales to explain
flat-rotation curves of galaxies \cite{mannheim} but the debate is
open and final tests should be achieved.

\subsection{ Newtonian limit in scalar-tensor gravity} \lb{sec:recovering}
Also in the case of scalar-tensor gravity, the Newtonian limit
gives non-standard gravitational potentials with exponential
contributions  directly depending on the parameters of scalar
field coupling and potential. Before starting with our analysis,
we need a choice for the up to now arbitrary functions $f(\p)$ and
$\v$ used in action (\ref{eq:high_scalar-tensor_action}). A rather
general choice is given by
 \beq\lb{410}
 f(\p)=\xi \p^m\,,
 \eeq
 \beq
 \lb{411} \v=\lam \p^n\,,
 \eeq
 where $\xi$ is a coupling constant, $\lambda$ gives the self--interaction
potential strength,  $m$ and $n$ are arbitrary parameters. This a
physically motivated choice, in agreement with the existence of a
Noether symmetry in the action
(\ref{eq:high_scalar-tensor_action}) as discussed in
\cite{cimento,cqg}. Furthermore, several scalar--tensor physical
theories (e.g. {\it induced gravity}) admit such a form for
$f(\phi)$ and $V(\phi)$.

It is instructive to develop in details the calculations in order
to see how extra contributions come out. In order to recover the
Newtonian limit, we write, as above, the metric tensor as
 \beq\lb{412}
 \gd=\eta\dmunu+h \dmunu\,,
 \eeq
 where $\eta\dmunu$ is the Minkoskwi metric and
$h\dmunu$ is a small correction to it. In the same way, we define
the scalar field $\psi$ as a perturbation, of the same order of
the components of $h\dmunu$, of the original field $\p$, that is
 \beq\lb{413}
 \p=\varphi_0+\psi\,,
 \eeq
where $\varphi_{0}$ is a constant of order unit. It is clear that
for $\varphi_{0}=1$ and $\psi=0$ Einstein GR is fully recovered.

To write in an appropriate form the Einstein tensor $G\dmunu$, we
define the auxiliary fields
 \beq\lb{414}
 \overline{h} \dmunu\equiv h\dmunu-\half\,\eta\dmunu h\,,
 \eeq
and
 \beq\lb{415}
 \sig\da\equiv {\overline h}_{\alp\beta,\gam}
 \eta^{\beta\gam}\,.
 \eeq
Given these definitions, to the first order in $h\dmunu$, we
obtain
 \beq\lb{416}
 G\dmunu=-\half \left\{\Box_{\eta} {\overline
 h}_{\mu\nu}+\eta\dmunu
 \sig_{\alp,\beta}\eta\uab-\sig_{\mu,\nu}-\sig_{\nu,\mu}
 \right\}\,,
 \eeq
 where $\Box_{\eta}\equiv \eta\umunu \pa_{\mu}
\pa_{\nu}$. We have not fixed the gauge yet.

We pass now to the right hand side of field equations
(\ref{eq:high_scalar-tensor_field_values}), namely to the
effective stress energy tensor. Up to the second order in $\psi$,
the coupling function $f(\p)$ and the potential $\v$, by using
Eqs.(\ref{410}) and (\ref{411}), become
 \beq\lb{417}
 f(\p)\simeq\xi \left(\varphi_0^m+m\varphi_0^{m-1}\,
 \psi+\frac{m(m-1)}{2}\varphi_0^{m-2}\,\psi^2\right)\,,
 \eeq
 \beq\lb{418}
 \v\simeq\lam \left(\varphi_0^{n}+n\,\varphi_0^{n-1}\psi+
 \frac{n(n-1)}{2}\,\varphi_0^{n-2}\psi^2\right)\,.
 \eeq

To the first order, the effective stress--energy tensor becomes
 \beq\lb{420}
 \tilde{T}\dmunu=-m \varphi_0^{2m-1}\,\eta\dmunu \Box_{\eta} \psi
 +m\, \varphi_0^{2m-1}\psi_{,\mu\nu}-\,\frac{\lambda\varphi_0^{m+n}}{2\xi}\eta\dmunu
 - (4\,\pi\tilde{G})\frac{\varphi_0^{m}}{\xi} \,T\dmunu\,,
 \eeq
 and then  the field equations are
 \beq\lb{421}
 \half \left\{\Box_{\eta} {\overline
 h}_{\mu\nu}+\eta\dmunu \sig_{\alp,\beta}\eta\uab-\sig_{\mu,\nu}-
 \sig_{\nu,\mu} \right\}=m\,\varphi_0^{2m-1} \eta\dmunu \Box_{\eta} \psi -
 m\,\varphi_0^{2m-1}\psi_{,\mu\nu}+
 \eeq
 $$
 +\,\frac{\lambda\varphi_0^{m+n}}{2\xi}\eta\dmunu +
 (4\,\pi\tilde{G})\frac{\varphi_0^{m}}{\xi}\,T\dmunu\,.
 $$
 We can eliminate the  term proportional to $\psi_{,\mu\nu}$  by choosing an
appropriate gauge. In fact, by writing the auxiliary field
$\sigma_{\alpha}$, given by Eq.(\ref{415}), as
 \beq\lb{422}
 \sig\da=m \,\varphi_0^{2m-1}\psi_{,\alp}\,,
 \eeq
 field equations (\ref{421}) read
 \beq\lb{423}
 \Box_{\eta} \overline {h}\dmunu-m\,\varphi_0^{2m-1} \eta\dmunu \Box_{\eta}
 \psi\simeq \frac{\lambda\varphi_0^{m+n}}{\xi}\,\eta_{\mu\nu}+
 (8\,\pi \tilde{G}) \frac{\varphi_0^{m}}{\xi}\, T\dmunu
 \eeq
 By defining the auxiliary field with components
$\tilde {h}\dmunu$ as
 \beq\lb{424}
 \tilde {h}\dmunu \equiv
 {\overline h}\dmunu -m \varphi_0^{2m-1}\eta\dmunu \psi\,,
 \eeq
 the field equations take the simpler form
 \beq\lb{425}
 \Box_{\eta} \tilde{h}\dmunu=\frac{\lambda\varphi_0^{m+n}}{\xi}\,\eta\dmunu+
  (8\,\pi \tilde{G})\frac{\varphi_0^{m}}{\xi}\, T\dmunu
 \eeq
 The original perturbation field $h\dmunu$ can
be written in terms of the new field as (with $\tilde{h}\equiv
\eta^{\mu\nu} \tilde{h}\dmunu$)
 \beq \lb{426}
 h\dmunu=\tilde{h}\dmunu-\half\,\eta\dmunu \tilde {h}-
 m \,\varphi_0^{2m-1}\eta\dmunu \psi\,.
 \eeq
We turn now to the Klein-Gordon Eq.(\ref{eq:scalar-tensor_KG_eq}).
$\Box_{\eta}\psi$ can be written, from the linearized Klein-
Gordon equation, in terms of the matter stress energy tensor and
of the potential term. If we calculate the scalar invariant of
curvature $R=\gu R\dmunu$ from Eq.(\ref{eq:T_eff_expression}), we
find
 \beq\lb{427}
 \Box\p+\frac{f(\p)'}{f(\p)} \left( \half\, g\uab \p\ddea \p\ddeb-2\,
 \v-3\,\Box f(\p)-4\,\pi \tilde{G} \,T\right)+ \vp=0\,,
 \eeq
 and, to the first order, it reads
 \beq\lb{428}
 \Box_{\eta}\psi+\frac{\lambda (n-2m)(n-1)\varphi_0^{n-2}}{1-3\xi m^2\varphi_0^{m-2}}\psi=
 \frac{\lambda (2m-n)\varphi_0^{n-1}}{1-3\,\xi m^2\varphi_0^{m-2}}+
  \frac{4\,\pi \tilde{G} m^2}{(1-3\,\xi m^2\varphi_0^{m-2})\varphi_0} T\,.
 \eeq
We work in the weak-field and slow motion limits, namely we assume
that the matter stress-energy tensor $T\dmunu$ is dominated by the
mass density term and we neglect time derivatives with respect to
the space derivatives, so that $\Box_{\eta}\rightarrow -\Delta$,
where $\Delta$ is the ordinary Laplacian operator in flat
spacetime. The linearized field equations (\ref{425}) and
(\ref{428}) have, for point--like distribution of matter\footnote{
To be precise, we can define a Schwarzschild mass of the form
$$M=\int(2T^{0}_{0}-T^{\mu}_{\mu})\sqrt{-g}d^3x \,.$$},
which is $\rho(r)=M\delta(r)$, the
following solutions:\\
for $n\neq 2m$, $n\neq 1$, we get
 \beqa
 h_{00}&\simeq & \left[(4\pi
 \tilde{G})\frac{\varphi_0^{m}}{\xi}\right]\frac{M}{r}-
 \left[\frac{4\pi\lambda\varphi_0^{m+n}}{\xi}\right]r^2-
 \left[(4\pi\tilde{G})\frac{m^2\varphi_0^{2m-2}M}{1-3\,\xi
m^2\varphi_0^{m-2}}\right]\frac{e^{-pr}}{r}
 +\nonumber \\
 & & -
 \left[\frac{4\pi m\varphi_0^{2m}}{n-1}\right]\cosh (pr)\,, \label{428a} \\
 h_{il} & \simeq & \delta_{il}\left\{\left[(4\pi \tilde{G})
\frac{\varphi_0^{m}}{\xi}\right]\frac{M}{r}+
 \left[\frac{4\pi\lambda\varphi_0^{m+n}}{\xi}\right]r^2+
 \left[(4\pi\tilde{G})\frac{m^2\varphi_0^{2m-2}M}{1-3\,\xi m^2
\varphi_0^{m-2}}\right]
 \frac{e^{-pr}}{r} \right\}+\nonumber \\
   & & -\delta_{il}\left[\frac{4\pi \varphi_0^{2m}m}{n-1}\right]\cosh (pr) \,, \label{
428b} \\
 \psi & \simeq &\left[ (4\pi\tilde{G})\, \frac{mM}{1-3\,\xi
 m^2\varphi_0}\right]\frac{e^{-pr}}{r}-\left[\frac{4\pi\varphi_0
}{n-1}\right]\cosh
 (pr)\,,\label{428c}
 \eeqa
 where the parameter $p$ is given by
 \beq\label{428d}
 p^2=\frac{\lambda (n-2m)(n-1)\varphi_0^{m-2}}{1-3\,\xi m^2\varphi_0^{m-2}}\,.
 \eeq
For $n=2m$, we obtain
 \beqa \lb{431}
 h_{00} & \simeq &\left[\frac{(4\pi\tilde{G})\varphi_0^{m}(1-4\,\xi m^2\varphi_0^{m-2})}{\xi
 (1-3\,\xi m^2\varphi_0^{m-2})}\right]\frac{M}{r}-\left[\frac{4\pi
\lambda\varphi_0^{m+n}}{\xi}\right]r^2
 -\Lambda\,,\\
 \lb{432}
 h_{il} & \simeq &
 \delta_{il}\left\{\left[\frac{(4\pi\tilde{G})\varphi_0^{m}(1-2\xi m^2\varphi_0^{m-2})}{\xi
 (1-3\,\xi m^2\varphi_0^{m-2})}\right]\frac{M}{r}+
 \left[\frac{4\pi\lambda\varphi_0^{m+n}}{\xi}\right]r^2+
\Lambda \right\} \,, \\
 \lb{433}
 \psi & \simeq & \left[\frac{(4\pi\tilde{G})m}{(1-3\,\xi
m^2\varphi_0^{m-2})\varphi_0}\right]
 \frac{M}{r}+\psi_0
 \eeqa
 where $\Lambda=m\varphi_0^{2m-1}\psi_0$ and $\psi_0$ are arbitrary
integration constants. Let us note that the values $m=1$, $n=2$
and $m=2$, $n=4$ correspond to the well known couplings and
potentials, i.e. $f\sim \phi$, $V\sim \phi^2$ and $f\sim \phi^2$,
$V\sim \phi^4$, respectively.

Finally for
 $n=1$, we obtain
 \beq \lb{433a}
 h_{00}  \simeq \left[\frac{(4\pi\tilde{G})\varphi_0^{m}(1-4\,\xi m^2\varphi_0^{m-2})}{\xi
 (1-3\,\xi m^2\varphi_0^{m-2})}\right]\frac{M}{r}-
\left[\frac{4\pi\lambda\varphi_0^{m+n}[1-\xi
m(m+1)\varphi_0^{m-2}]}
 {\xi(1-3\xi m^2\varphi_0^{m-2})}\right]r^2,
\eeq \beq
 \lb{433b}
 h_{il}  \simeq
 \delta_{il}\left\{\left[\frac{(4\pi\tilde{G})\varphi_0^{m}(1-2\,\xi m^2\varphi_0^{m-2})}{\xi
 (1-3\xi m^2\varphi_0^{m-2})}\right]\frac{M}{r}+
\left[ \frac{4\pi\lambda\varphi_0^{m+n}
 [1-\xi m(m+1)\varphi_0^{m-2}]}{\xi(1-3\xi
m^2\varphi_0^{m-2})}\right]r^2\right\}, \eeq \beq
 \lb{433c}
 \psi  \simeq \left[\frac{(4\pi\tilde{G})m}{(1-3\xi
 m^2)\varphi_0}\right]\frac{M}{r}+
 \left[\frac{4\pi\lambda (2m-1)\varphi_0^{n-1}}{1-3\xi m^2\varphi_0^{m-2}}
\right] r^2 \,.
 \eeq
If we demand the $(0,0)$--component of the field Eq.(\ref{425}),
when $\lam=0$, to read as the usual Poisson equation (that is
nothing else but a definition of the mass)
 \beq \lb{434}
 \Delta\Phi=4\,\pi  G_{N} \rho\,,
 \eeq
 where $\Phi$, linked with the metric tensor by the relation
$h_{00}=2\,\Phi$, is the Newtonian potential, we have to put
 \beq\lb{435}
 G_{N}=-\frac{\varphi_0^{m}}{2\,\xi}\left(\frac{1-4\,\xi m^2\varphi_0^{m-2}}
 {1-3\,\xi m^2\varphi_0^{m-2}}\right) \tilde {G}\,.
 \eeq
 We may now rewrite the nonzero components of $h\dmunu$ and the
scalar perturbed field. Let us take into account, for example,
Eqs. (\ref{431})--(\ref{433}). We get
 \beq\lb{436}
 h_{00}\simeq -\frac{2\,G_{N} M}{r}- \frac{4\pi\lambda\varphi_0^{m+n}}{\xi}\,r^2
 -\varphi_0^{2m-1}\psi_0
 \eeq
 \beq\lb{437}
 h_{il}\simeq \delta_{il}\left\{-\frac{2\,G_{N}M}{r}\left(\frac{1-2\,\xi  m^2\varphi_0^{m-2}}
 {1-4\,\xi m^2\varphi_0^{m-2}}\right)+\frac{4\pi\lambda\varphi_0^{m+n}}{\xi}\,r^2
 +m\varphi_0^{2m-1}\psi_0\right\} \,,
 \eeq
 and
 \beq\lb{438}
 \psi=-\frac{2\,G_{N} M}{r} \left( \frac{\xi m\varphi_0^{-m-1}}{1-4\,\xi
 m^2}\right) +\psi_0\,,
 \eeq
where the Newton constant explicitly appears. Similar
considerations hold in the other cases.

What we have obtained are solutions of the linearized field
equations, starting from the action of a scalar field nonminimally
coupled to the geometry, and minimally coupled to the ordinary
matter. Such solutions  depend on the parameters which
characterize the theory:  $\xi, m,n,\lambda$. The results of
Einstein GR are obtained for $f=f_{0}$, with $f_{0}$
negatively--defined due to the sign choice in the action
(\ref{eq:high_scalar-tensor_action}). As we can easily see from
above, in particular from Eqs.(\ref{436})--(\ref{438}), we have
the usual Newtonian potential and a sort of {\it cosmological
term} ruled by $\lambda$ which, from the Poisson equation, gives a
quadratic contribution.

We consider now the Brans-Dicke-like action (\ref{eq:B-D_action})
where $\omega=\omega\left(\pv\right)$. It is actually simple to
see, from the  Eqs.(\ref{eq:T_eff_new}) and (\ref{eq:KGeqs}),
that, if we want to limit ourselves to the linear approximation,
we may consider as well $\omega=$constant. The link with the
results that follows from the action (\ref{eq:B-D_action}) are
given by the transformation laws (\ref{eq:choice}), that is
 \beq \lb{439}
 \omega\left(\pv(\p)\right)=\frac{1}{2\,\xi m^2} \, \p^{2- m}\,.
 \eeq
 The potential term $W\left(\pv\right)$ in the linear
approximation, behaves as $V(\phi)$. Results in the approaches
given by the actions (\ref{eq:high_scalar-tensor_action}) and
(\ref{eq:B-D_action}) are completely equivalent.

\subsection{An application: Newtonian limit of String-Dilaton Gravity}
The above approach can be applied to the case of string-dilaton
gravity which is a remarkable example of scalar-tensor theory
which have to be investigated also in the Newtonian limit.

Let us shortly outline why such a theory is so interesting from
cosmological and astrophysical viewpoints.

String-dilaton gravity seems to yield one of the most promising
scenarios in order to solve several shortcomings of standard and
inflationary cosmology \cite{vafa}. First of all, it addresses the
problem of the initial singularity, which is elegantly solved by
invoking a maximal spacetime curvature directly related to the
string size \cite{gasperini}. Besides, it introduces a wide family
of cosmological solutions which comes out thanks to the existence
of a peculiar symmetry called {\sl duality}, which holds at string
fundamental scales, as well as at cosmological scales
\cite{veneziano}. In practice, if $a(t)$ is a cosmological
solution of a string--dilaton model, also $a^{-1}(t)$ has to be
one by a time reversal $t\ra-t$. In this case, one can study the
evolution of the universe toward $t\ra+\infty$ as well as toward
$t\ra-\infty$. The junction of these two classes of solutions, at
some maximal value of curvature, considered as branches of a (up
to now not found) general solutions, has to be in agreement with
inflationary paradigm and solve the initial values and singularity
problems of standard
cosmological model \cite{gasperini}.\\
 String-dilaton cosmological models
 come out from the low--energy limit of (super)string theory which
can be considered as one of the most serious attempt to get the
great unification of all forces in Nature in the last thirty
years. This theory avoids the shortcomings of quantum field
theories due, essentially, to the point--like nature of particles
-- the renormalization -- and includes gravity in the same
conceptual scheme as the other fundamental interactions -- the
graviton is just a string {\it mode} like
the other gauge bosons \cite{schwarz}.\\
\in However, despite the quality and the variety of theoretical
results, the possibility to test experimentally the full
predictions of the theory is not yet available. The main reason
for this failure is that the Planck scale -- $\sim~10^{19}$ GeV
--, where the string effects become relevant, is  unreachable in
today's high--energy physics experiments.
 Therefore, cosmology and astrophysics remain the only open ways for observational investigations and
 detecting {\it remnants}
 of primordial processes could lead to test this theory.
 Furthermore, a lot of open questions of astrophysics, like dark matter,
 relic gravitational-wave background, large-scale structure, primordial
 magnetic fields and so on could be explained by strings and their
 dynamics.\\
The key element of string--dilaton gravity \cite{vafa,tseytlin},
 in low--energy limit, is the fact that a dynamics, consistent with duality,
 can be implemented only by taking into account massless modes (zero modes) where the scalar mode
 (the dilaton) is non-minimally coupled to the other fields. The tree--level effective action, in general,
 contains a second--rank symmetric tensor field (the metric), a scalar field (the dilaton)
 and a second--rank antisymmetric tensor field (the so--called Kalb--Ramond {\it universal} axion).
 Such an action can be recast as a scalar--tensor theory, \eg induced gravity, where the gravitational
 coupling is a function of the dilaton field \cite{nmc,cqg,vafa}. Then, it is legitimate to study the
 Newtonian limit of the string--dilaton gravity to see what is its behavior in the weak--field and
 slow--motion approximations. This approach is useful to investigate how string--dilaton dynamics
 could affect scales shorter than cosmological ones. The issue is to search for effects of the coupling
 and the self--interaction potential of the dilaton at scales of the order of Solar System or Galactic size.\\
\in Calculations can be developed in the string frame since one
wants to understand the role of a dilaton--non-minimal coupling in
the recovering of the Newtonian limit. In the string-dilaton
theory, it is necessary to introduce a scalar potential with a
power law expression of the form ${\displaystyle \v=\lam~\p^n}$ (see \cite{caplamb} for details).\\
\in The tree--level string--dilaton effective action, the lowest
order in loop expansion, containing all the massless modes,
without higher--order curvature corrections (\ie without the
Gauss--Bonnet invariant) is

\beqa
 {\s}&=&-\frac{1}{2 \, \lam_s^{d-1}}\int d^{d+1}x \, \sqrt{-g}e^{-\p} \,
 \left[R+(\nabla\p)^2 - \frac{1}{12} \, H_{\mu\nu\alp} \, H^{\mu\nu\alp}+ \v \right]+ \no \\
 & &+\int d^{d+1}x \, \sqrt{-g} \, {\l}_m
\lb{eq:string-dilaton_action} \eeqa

\noin where $R$ is the Ricci scalar, $\p$ the dilaton field, $\v$
the dilaton self--interaction potential.
$H_{\mu\nu\alp}=\pa_{[\mu}B_{\nu\alp]}$ is the full antisymmetric
derivative of the Kalb--Ramond axion tensor and ${\l}_{m}$ is the
Lagrangian density of other generic matter sources. The theory is
formulated in $(d+1)$--dimensions, $d$ being the number of spatial
dimensions, and $\lam_s$  the string fundamental minimal length
parameter. To the lowest order, the effective gravitational
coupling is given by

\beq
 \sqrt{8 \, \pi \, G_{N}} \; = \; \lam_{p} \; = \; \lam_{s} \, e^{\p/2}
 \lb{eq:thecoupling}
\eeq

\noin where $G_N$ is the Newton constant and $\lam_{p}$ is the Planck length.
Units have been chosen such that $2\lam_{s}^{d-1}=1$ so that $e^\p$ is the
$(d+1)$--dimensional gravitational coupling.\\
\in Clearly, the string-dilaton effective action
(\ref{eq:string-dilaton_action}), apart the Kalb-Ramond term, is a
particular case of the general scalar-tensor action
(\ref{eq:high_scalar-tensor_action}) where the transformations
\beq \p\,\longrightarrow\,\exp[-\p]\,,\qquad
f(\p)\,\longrightarrow \frac{1}{2}\exp[-\p]\,,\qquad
V(\p)\,\longrightarrow V(\p)\exp[-\p]\,,\eeq are performed.

\in The field equations are derived by varying the action
(\ref{eq:string-dilaton_action}) with respect to $\gd$, $\p$, and
$B\dmunu$. One gets, respectively,

\beqa
 \Gd + \, \nabla\dmu \, \nabla\dnu \p \, + \, \half \,\gd \, \left[(\nabla\p)^2-2\,\nabla^2\p-\v+\frac{1}{12} \,
 H_{\alp\beta\gam} \, H^{\alp\beta\gam}\right] \no \\
 -\frac{1}{4} \, H_{\mu\alp\beta} \,H\dnu^{\phantom{\nu}\alp\beta} \; = \; \frac{e^{\p}}{2} \,T\dmunu
 \lb{eq:variation_1} \\
 R+2 \,\nabla^2 \,\p-(\nabla \p)^2 +V-V'-\frac{1}{12} \,H_{\mu\nu\alp} \,H^{\mu\nu\alp} \; = \; 0
 \lb{eq:variation_2} \\
 \nabla \dmu \left(e^{-\p}H^{\mu\alp\beta}\right) \; = \; 0 \lb{eq:variation_3}
\eeqa

\noin where, as above, $\Gd$ is the Einstein tensor,
$T \dmunu$ the stress--energy tensor of matter sources and $V'=dV/d\p$.\\
\in Let us assume that standard matter is a perfect fluid
minimally coupled to the dilaton. Otherwise, in the non-minimally
coupled case, a further term should be taken into account in
Eq.~(\ref{eq:variation_2}). The above equations  form a system of
tensor equations in $(d+1)$
 dimensions which assigns the dynamics of $\gd$, $\phi$, and $B\dmunu$.\\
\in Now let us consider the weak field approximation in order to derive the PPN limit of the theory.
Obviously, all the invariants of the full theory are not preserved if equations are linearized.
For example, duality is lost in the linearized solutions. However, this is not a problem in the
present context since there is the assumption of a regime well--separated from early singularity
where duality is adopted to solve cosmological shortcomings.
Here, the aim is to check if {\it remnants} of primordial string--dilaton dynamics are detectable
at our nearest scales (Solar System or Galaxy).\\
\in Moreover, the weak field approximation for the axion gives
only second order terms with respect to $\hd$ and $\psi$ in the
field equations, so that its contribution in the following
considerations can be discarded. A physical interpretation of this
fact could be that the production of primordial magnetic fields,
considered as ``seeds'' for the today's observed large magnetic
fields of galaxies
\cite{giovannini} is a second order effect if it is due to $H_{\mu\nu\alp}$.\\
\in Using  the approximated expressions (\ref{412}) and
(\ref{413}) for the fields and the natural assumption
${\displaystyle \v=\lam~\p^n}$ for the scalar field potential, the
field equations (\ref{eq:variation_1}) and (\ref{eq:variation_2})
become respectively

\beqa
  \Box_{\eta} \bar{h}\dmunu +\eta \dmunu \, \sig_{\alp,}^{\phantom{,\alp}\alp}-
 (\sig_{\mu,\nu}+\sig_{\nu,\mu})-2 \, \psi_{,\,\mu\nu}+2 \, \eta \dmunu \,
 \Box_{\eta}\psi+\lam\,(\eta \dmunu+\hd)\,\pv_0^n+ \no \\ \lb{eq:Ein_new}
  +\lam\, n\,\pv_0^{n-1} \,\eta \dmunu \, \psi \;\simeq\; -e^{\pv_0}\, T \dmunu \\
  2 \, \Box_{\eta} \,\psi+\half \,\Box_{\eta} \bar{h}
  +\sig_{\alp ,}^{\phantom{\alp}\alp} +\lam\,\pv_0^n[n\,\pv_0^{-1}-n\,(n-1)\,\pv_0^{-2}]\,\psi
  +\lam \,\pv_0^n[1-n \, \pv_0^{-1}]=0 \; .
\lb{eq:KG_new} \eeqa

\in The field equation (\ref{eq:variation_3}) has been discarded
since it gives only terms higher than the linear order. The term
proportional to $\psi_{,\mu\nu}$ can be eliminated by choosing an
appropriate gauge. In fact, by writing the auxiliary field
$\sig_{\alp}$ as

\beq
 \sig_{\mu}=-\psi_{,\,\mu}
\lb{eq:gauge} \eeq

\in Eq.~(\ref{eq:Ein_new}) reads

\beq
 \Box_{\eta} \bar{h} \dmunu + \eta \dmunu \Box_{\eta} \psi
 +\lam\,\pv_0^n\left[\eta \dmunu \left(1+\frac{n}{\pv_0}\psi\right)+ \hd\right] \;\simeq\; - e^{\pv_0}\, T \dmunu \; .
\lb{eq:eq_K-G} \eeq

\in In this approximation, the terms in $\hd$ and $\psi$ in the
brackets can be neglected as $\hd\ll\eta \dmunu$ and $\psi\ll 1$.
Eq.~(\ref{eq:eq_K-G}) then becomes

\beq
 \Box_{\eta} \bar{h} \dmunu+\eta \dmunu \,\Box_{\eta} \,\psi
 +\lam\,\pv_0^n\,\eta \dmunu \;\simeq\; - e^{\pv_0} \, T \dmunu \; .
\lb{eq:eq_K-G_1} \eeq

\in By defining the further auxiliary field

\beq
 \tilde{h} \dmunu \;\equiv\; \bar{h} \dmunu + \eta \dmunu \, \psi
\lb{eq:aux_field} \eeq

\noin one gets the final form

\beq
 \Box_{\eta} \tilde{h} \dmunu+\lam\,\pv_0^n\,\eta \dmunu \;\simeq\; -e^{-\pv_0}\, T\dmunu \; .
\lb{eq:eq_K-G_2} \eeq

\in The original perturbation field $\hd$ can be written in terms
of the new field, \ie

\beq
 \hd \;=\; \tilde{h} \dmunu-\half\,\eta \dmunu \, \tilde {h}+\eta \dmunu \,\psi
\lb{eq:aux_field_1} \eeq

\noin with $\tilde{h}\equiv \eta \umunu \tilde{h} \dmunu$.\\
\in Let us turn now to the Klein-Gordon equation
(\ref{eq:KG_new}). With some algebra, it can be recast in the form

\beq
  \left(\Box_{\eta}+c_1^2\right) \, \psi \;\simeq\; -\frac{e^{-\pv_0}}{2} \, T-\Phi_0
\lb{eq:KG_new_3} \eeq

\noin where $T$ is the trace of the stress-energy tensor of
standard matter and the constants are

\beq
 c_1^2 \;=\; \lam \, \pv_0^{n} \, \left[n(n-1) \, \pv_0^{-2}-n \, \pv_0^{-1}\right]\,,\qquad
\Phi_0 \;=\; \left(3-n \,\pv_0^{-1}\right) \,\lam \,\pv_0^{n} \; .
\lb{eq:constant_dilaton} \eeq

\in The linearized field equations (\ref{eq:eq_K-G_2}) and
(\ref{eq:KG_new_3}) have, for point--like distribution of matter,
the following solutions

\beqa
 h_{00} \;\simeq\; -\frac{2 \,G_N \,M}{r}(1-e^{-c_{1} \,r})+c_2 \,r^2+c_3 \, \cosh (c_1r) \lb{eq:dilaton_1} \\
 h_{ii} \;\simeq\; -\frac{2 \,G_N \,M}{r}(1+e^{-c_{1} \,r})-c_2 \, r^2-c_3 \, \cosh (c_1r) \lb{eq:dilaton_2} \\
 \psi \;\simeq\; \frac{2 \,G_N \,M}{r}e^{-c_{1} \,r}+c_3 \, \cosh (c_1 \, r) \lb{eq:dilaton_3}
\eeqa

\noin where the coefficients $c_2$ and $c_3$ are

\beq
 c_2 \; = \; 2 \, \pi \, \lam \, \pv_0^n \,,\qquad c_3 \; = \; \frac{3-n \, \pv_0^{-1}}{n(n-1) \, \pv_0^{-2}-n \, \pv_0^{-1}}
\lb{eq:constant_dilaton_1} \eeq

\in Here the functions (\ref{eq:dilaton_1}) and (\ref{eq:dilaton_2}) are the non-zero components of
$\hd$ while Eq.~(\ref{eq:dilaton_3}) is the perturbation of the dilaton.
Standard units have been restored.\\
\in When $\lam=0$ and in the slow motion limit, the
$(0,0)$--component of the field Eq.~(\ref{eq:eq_K-G_2}) reduces to
the  Poisson equation (\ref{434}).

\in Solutions of the linearized field equations have been
obtained, starting from the action of a scalar field non-minimally
coupled to the geometry, and minimally coupled to the ordinary
matter (specifically the string-dilaton gravity). Such solutions
explicitly depend on the parameters $\pv_0,n,\lam$ which assign
the model in the class of
actions~(\ref{eq:string-dilaton_action}). At this point,
string-dilaton gravity is completely equivalent to any
scalar-tensor theory of gravity and the physical considerations
are the same.

\section{\bf Discussion and Conclusions}

As a general remark, the Newtonian limit of ETG strictly depends
on the parameters of the scalar-field couplings, the
self--interaction potentials and the higher-order coupling terms
as it has to be by a straightforward PPN parametrization
\cite{will}.

As we have seen,  the self--interactions (scalar or gravitational)
give rise, in any case, to corrections to the Newtonian potential.
Such corrections are, in general,  constant, quadratic or
Yukawa--like terms. Due to these facts,  ETG could be the
potential natural candidates   to solve several astrophysical
problems as the flat rotation curves of spiral galaxies
\cite{stelle,mannheim,kenmoku} (and then the up to now not
revealed presence of consistent bulks of dark matter), the
apparent, anomalous long-range acceleration revealed by several
spacecrafts \cite{anderson} . Essentially the corrections depends
on the strength of the couplings and the ``masses" of the scalar
fields (where, also higher-order curvature terms can be considered
under the standad of "scalar fields" using conformal
transformations). It is worth noting that we get always scale
lengths where Einstein GR can be, at least approximately,
recovered. This fact could account why measurements {\it inside}
the Solar System confirm such theory while {\it outside} of it
there are probable deviations \cite{anderson}.

\in We can make these considerations more precise discussing some
of the PPN solutions which we found above. For example, let us
take into account the Newtonian-like potential,  derived in PPN
approximation, from string-dilaton gravity.\\
\in As shown, the solution given in Eq.~(\ref{eq:dilaton_1}) can
be read as a Newtonian potential with exponential and quadratic
corrections, \ie

\beq
 \Phi(r) \;\simeq\; -\frac{G_N \, M}{r} \, (1-e^{-c_{1} \,r})+\frac{c_2}{2} \, r^2+\frac{c_3}{2} \, \cosh (c_1 \, r) \; .
\lb{eq:grav_const}
 \eeq

\in In general,  most of ETG  have a weak field limit of this form
(see also  \cite{schmidt1,stelle,kenmoku}), that is

\beq
 \Phi(r) \;=\; -\frac{G_N \, M}{r} \, \left[1+\sum_{k=1}^{n} \, \alp_k \, e^{-r/r_k}\right]
\lb{eq:yukawa} \eeq

\noin where $G_{N}$ is the value of the gravitational constant as
measured at infinity and $r_k$ is the interaction length of the
$k$-th component of the non-Newtonian corrections. The amplitude
$\alp_k$ of each component is normalized to the standard Newtonian
term and the signs of the $\alp_k$ coefficients indicate if the
corrections are attractive or repulsive \cite{will}. Besides, the
variation of the gravitational coupling is included. As an
example, let us take into account only the first term of the
series in Eq.~(\ref{eq:yukawa}) which is usually considered as the
leading term (this choice is not sufficient if other corrections
are needed). One has

\beq
 \Phi(r) \;=\; -\frac{G_N \, M}{r} \,\left[1+\alp_1 \, e^{-r/r_1}\right]\,.
\lb{eq:yukawa1} \eeq

\in The effect of non-Newtonian term can be parameterized by
$(\alp_1,\,r_1)$. For large distances, at which $r\gg r_1$, the
exponential term vanishes and the gravitational coupling is
$G_{N}$. If $r\ll r_1$, the exponential becomes of order unity; by
differentiating Eq.~(\ref{eq:yukawa1}) and comparing with the
gravitational force measured in laboratory, one gets

\beq
 G_{lab} \;=\; G_{N} \, \left[1+\alp_1 \,\left(1+\frac{r}{r_1}\right)e^{-r/r_1}\right]
\simeq G_{N} \, (1+\alp_1) \lb{eq:yukawa2} \eeq

\noin where $G_{lab}=6.67\times 10^{-11}$ Kg$^{-1}$m$^3$s$^{-2}$
is the usual Newton constant measured by the most recent
Cavendish-like experiments. Of course, $G_{N}$ and $G_{lab}$
coincide in the standard gravity. It is worthwhile noticing that,
asymptotically, the inverse square law holds but that the measured
coupling constant differs by a factor $(1+\alpha_1)$. In general,
any  correction introduces a characteristic length that acts at a
certain scale for the self-gravitating systems. The range of $r_k$
of the $k$th-component of non-Newtonian force can be identified
with the mass $m_k$ of a pseudo-particle whose Compton's length is

\beq
 r_k \;=\; \frac{\hbar}{m_k \, c} \; .
\lb{eq:compton} \eeq

\in This means that, in the weak energy limit, fundamental
theories which attempt to unify gravity with the other forces
introduce, in addition to the massless graviton, particles {\it
with mass} which carry the gravitational force \cite{gibbons}.

\in There have been several attempts to constrain $r_k$ and
$\alp_k$ (and then $m_k$) by experiments on scales in the range $1
\,\mbox{cm}<r< 1000\, \mbox{km}$, using totally different
techniques \cite{fischbach,speake,eckhardt1}. The expected masses
for particles which should carry the additional gravitational
force are in the range $10^{-13} \mbox{eV}<m_k< 10^{-5}\,
\mbox{eV}$. The general outcome of these experiments, even
retaining only the term $k=1$, is that a ``geophysical window''
between the laboratory and the astronomical scales has to be taken
into account. In fact, the range

\beq |\alp_1|\sim 10^{-2}\,,\qquad r_1 \;\sim\; 10^2\div 10^3
\,\mbox{m} \lb{eq:range} \eeq

\noin is not excluded at all in this window by current measurements.
 An interesting suggestion has been made by Fujii \cite{fujii1},
 which proposed that the exponential deviation from the Newtonian standard potential
 (the ``fifth force'') could arise from the microscopic
 interaction which couples to nuclear isospin and baryon number.\\
\in The astrophysical counterpart of these non-Newtonian
corrections seemed ruled out as recent experimental tests of GR
``exactly'' predict the Newtonian potential in the weak energy
limit, inside the Solar System. Besides, indications of an
anomalous, long--range acceleration have been found during the
data analysis of Pioneer 10/11, Galileo, and Ulysses spacecrafts
(which are now almost outside the Solar System) and made these
Yukawa--like corrections come again into play \cite{anderson}.
Furthermore, Sanders \cite{sanders} reproduced the flat rotation
curves of spiral galaxies by using

\beq \alp_1 \;=\; -0.92\,,\qquad r_1\sim 40\,\mbox{kpc} \; .
\lb{sand} \eeq

\in His main hypothesis is that the additional gravitational interaction
is carried by an ultra-soft boson whose range of mass is $m_1\sim 10^{-27}\div 10^{-28}$eV.
The action of this boson becomes efficient at galactic scales without the request of
enormous amounts of dark matter to stabilize the systems.
In addition, Eckhardt \cite{eckhardt1} uses a combination of two exponential
terms and gives a detailed explanation of the kinematics of galaxies and galaxy clusters,
without dark matter models, using arguments similar to those of Sanders.\\
\in It is worthwhile noticing that both the spacecrafts and galactic rotation curves
indications are outside the usual Solar System boundaries used to test GR. However,
the above authors do not start from any fundamental theory in order to explain the outcome
of Yukawa corrections. In their analysis, Yukawa terms are purely phenomenological.\\
\in Another important remark in this direction deserves the fact
that some authors \cite{mcgaugh} interpret the recent data coming
from cosmic microwave background experiments, BOOMERANG
\cite{debernardis} and WMAP \cite{WMAP}, in the framework of
{\it modified Newtonian dynamics}, without invoking any dark matter model.\\
\in All these facts point toward the line of thinking that ``corrections''
to the standard gravity have to be seriously taken into account. Let us now
 turn back to the above solutions (\ref{eq:dilaton_1})--(\ref{eq:dilaton_3}),
 in particular to the gravitational potential (\ref{eq:grav_const}).
 It comes out from the weak field limit (PPN approximation) of a string-dilaton effective action
 (\ref{eq:string-dilaton_action}). The specific model is
 singled out by the number of spatial dimension $d$ and the
  form of the self-interaction potential $\v$ where the quite general class $\v=\lam\phi^n$ has been assumed.\\
\in Without loosing of generality, one can assume $\varphi_0=1$.
This means that for $\phi=1$ the standard gravitational coupling
is restored in the action (\ref{eq:string-dilaton_action}).
However, the condition $\psi\ll 1$ must hold. For the choice $n=3$
in Eq.(\ref{eq:constant_dilaton_1}), one has

\beq
 \Phi(r) \;\simeq\;
 -\frac{G_N \, M}{r}(1-e^{-c_{1} \, r})+\frac{c_2}{2} \, r^2
\lb{eq:grav_const_1} \eeq

\noin where, beside the standard Newtonian potential, two
correction terms are present. Due to the definition of the
constants $c_{1,2}$ their strength directly depends on the
coupling $\lam$ of the self-interaction potential $\v$. If
$\lam>0$, one finds, from Eq.~(\ref{eq:constant_dilaton}) and
 Eq.~(\ref{eq:constant_dilaton_1}), that the first correction is a {\it repulsive}
 Yukawa-like term with $\alp_1=-1$ and $r_1=c_1^{-1}=\lam^{-1/2}$. The second correction
 is given by a kind of positive-defined cosmological constant $c_2$ which acts as a
  repulsive force\footnote{If in the Poisson equation there is a positive constant density,
   it gives rise to a repulsive quadratic potential in $r$ and then to a linear force.
   A positive constant density can be easily interpreted as a  cosmological constant.}
   proportional to $r$. On the other hand, if $\lam<0$, the first correction is oscillatory while
   the second is attractive.\\
\in From the astrophysical point of view, the first situation is
more interesting. Assuming that the dilaton is an ultra-soft boson
which
 carries the scalar mode of gravitational field, one gets
 that the length scale $\sim 10^{22}\div 10^{23}$ cm, is obtained
 if its mass range is $m\sim 10^{-27}\div 10^{-28}$eV.
 This length scale is necessary to explain the flat rotation curves of spiral galaxy.
 The second correction to the Newtonian potential can contribute to
 stabilize the system  being repulsive and acting as a constant
 density -- a sort of cosmological constant at galactic scales.  In general, if $\alp_1\sim-1$ the flat
 rotation curves of galaxies can be reconstructed \cite{sanders}. If the mass of the dilaton
 is in the range $10^{-13} \mbox{eV}<m<10^{-5} \mbox{eV}$ the ``geophysical windows''
 could also be of interest for experiments. Finally the mass range $m\sim 10^{-9}\div 10^{-10}$eV
 could be interesting at Solar System scales (for the allowed mass windows
 in cosmology see Ref.~\cite{gasperini2}). Similar analysis can be performed also for
 other values of $n$ corresponding to different models of the class
 defined in Eq.~(\ref{eq:string-dilaton_action}).\\
\in In conclusion, the weak energy limit of string-dilaton gravity
has been derived, showing that in this case the Newtonian
gravitational potential is corrected by exponential and quadratic
terms. These terms introduce natural length scales which can be
connected to the mass of the dilaton. If the dilaton is an
ultra-soft boson, one can expect observable effects at
astrophysical scales. If it is more massive, the effects could be
interesting at geophysical or microscopic scales.

This is just an example, but a similar analysis can be developed
also for other ETG and the clues are that wide classes of
gravitational phenomena, at microscopic, macroscopic and
astrophysical scales, could be interpreted by enlarging the scheme
of GR.


\begin{thebibliography}{99}
\bibitem{guth} A. Guth, \pr {\bf D 23}, 347 (1981).
\bibitem{weinberg} S. Weinberg, {\it Gravitation and Cosmology}, Wiley, 1972
                   New York N.Y.
\bibitem{brans} C. Brans and R.H. Dicke, \pr {\bf 124}, 925
(1961).
\bibitem{starobinsky} A.A. Starobinsky, \pl {\bf 91B} (1980) 99.
\bibitem{la} D. La and P.J. Steinhardt, \prl {\bf 62} (1989) 376.
\bibitem{vilenkin} A. Vilenkin, \pr {\bf 32 D} (1985) 2511.\\
E. Carugno, S. Capozziello, F. Occhionero, \pr {\bf D 47}(1993)
4261.
\bibitem{odintsov} I.L. Buchbinder, S.D. Odintsov, and I.L. Shapiro,
                   {\it Effective Action in Quantum Gravity}, IOP Publishing
                   (1992) Bristol.
\bibitem{will} C.M. Will, {\it Theory and Experiments in Gravitational
               Physics} (1993) Cambridge Univ. Press, Cambridge.
\bibitem{stelle} K. Stelle, \grg {\bf 9} (1978) 353.
\bibitem{sanders} R.H. Sanders, \araa {\bf 2} (1990) 1.
\bibitem{mannheim} P.D. Mannheim and D. Kazanas, \apj {\bf 342} (1989) 635.
                   O.V. Barabash and Yu. V. Shtanov, \pr {\bf 60 D}
                   (1999) 064008.
\bibitem{anderson} J.D. Anderson et al., \prl {\bf 81} (1998)
2858.\\ J.D. Anderson, et al., \pr {\bf D 65} (2002) 082004.
\bibitem{schmidt} I. Quant and H.-J. Schmidt, {\it Astron. Nachr.} {\bf 312}
                  (1991) 97.
\bibitem{ehlers} P. Schneider, J. Ehlers, and E.E. Falco, {\it Gravitational
                Lenses} Springer--Verlag (1992) Berlin.
\bibitem{krauss} L.M. Krauss and M. White, \apj {\bf 397} (1992) 357.
\bibitem{nottale} L. Nottale in {\it Dark Matter (Moriond Astrophysics
                  Meetings)}, J. Andouze and J. Tran Thanh Van eds.
                  (1988) Frontieres, Gif--sur--Yvette.
\bibitem{urbino} Ph. Tourrenc, {\it General Relativity and Gravitational
Waves} in Proc. of the Int. Summer School on Experimental Physics
of Grav. Waves, September 6-18, 1999, Urbino (Italy), Eds. M.
Barone et al., World Scientific (Singapore) 2000.
\bibitem{microSCOPE} http://www.onera.fr/prix-en/brun2002/edmond-brun2002.html
\bibitem{miniSTEP} http://astro.estec.esa.nl/astrogen/COSPAR/step/
\bibitem{rebka} R.V. Pound and G.A. Rebka, \prl {\bf 4}, 337
(1960).
\bibitem{HT} R.A. Hulse and J.H. Taylor, \apj {\bf 195}, L51
(1975).
\bibitem{AEGR} A. Einstein, {\it Annalen der Physik}, {\bf 49},
769 (1916).
\bibitem{eddington} A.S. Eddington, {\it The Observatory}, {\bf
42}, 119 (1919).
\bibitem{wald} R. Wald,  {\it General Relativity}, Chicago Univ.
Press, Chicago (1984).
\bibitem{landau} L.D. Landau and E.M. Lifshitz, {\it The Classical Theory of
Fields}, Pergamon, Oxford (1962).
\bibitem{MTW} C.W. Misner, K.S. Thorne, and J.A. Wheeler, {\it
Gravitation}, W.H. Freeman, San Francisco (1973).
\bibitem{thorne} K.S. Thorne,  {\it The theory of gravitational radion: an introductory
review}, in {\it Gravitational Radiation}, pp 1-57, ed. N.
Deruelle and T. Piran, North-Holland, Amsterdam (1983).
\bibitem{robertson} H.P. Robertson, \rmp {\bf 21}, 378 (1949).
\bibitem{MS} R. Mansouri and R.U. Sexl, \grg {\bf 8}, 497 (1977).
\bibitem{VS} I. Vetharanian and G.E. Stedman, \cqg {\bf 11}, 1069
(1994).
\bibitem{tourrenc} Ph. Tourrenc, T. Mellitti et al., \grg {\bf 28},
1071 (1996).
\bibitem{LL} A.P. Lightman and D.L. Lee, \pr {\bf D 8}, 364
(1973).
\bibitem{blanchet} L. Blanchet, \prl {\bf 69}, 559 (1992).
\bibitem{DP} T. Damour and A.M. Polyakov \np {\bf B 423}, 532
(1994).
\bibitem{lator} S.G. Turyshev, M. Shao and K. Nordtvedt Jr.,
invited talk presented at {\it AIAA Meeting Space 2003}, September
23, Long Beach (CA).
\bibitem{francaviglia} G. Magnano, M. Ferraris, and M.
Francaviglia, \grg {\bf 19}, 465 (1987).
\bibitem{ottewill} J. Barrow and A.C. Ottewill, {\it J. Phys. A: Math.
Gen.} {\bf 16}, 2757 (1983).
\bibitem{birrell} N.D. Birrell and P.C.W. Davies, {\it Quantum Fields in Curved
Space}, Cambridge Univ. Press, Cambridge (1982).
\bibitem{vilkovisky} G. Vilkovisky, \cqg {\bf 9}, 895 (1992).
\bibitem{veneziano} M. Gasperini and G. Veneziano, \pl {\bf 277B}
(1992) 256.
\bibitem{bondi} H. Bondi, {\it Cosmology}, Cambridge Univ. Press,
Cambridge (1952).
\bibitem{cimento} S. Capozziello, R. de Ritis, C. Rubano, and P.
Scudellaro, {\it La Rivista del Nuovo Cimento} {\bf 4} (1996) 1.
\bibitem{sciama} D.W. Sciama, \mnras {\bf 113}, 34 (1953).
\bibitem{teyssandier} P. Teyssandier and Ph. Tourrenc, {\it J. Math.
Phys.} {\bf 24}, 2793 (1983).
\bibitem{maeda} K. Maeda, \pr {\bf D 39}, 3159 (1989).
\bibitem{wands} D. Wands, \cqg {\bf 11}, 269 (1994).
\bibitem{conf} S. Capozziello, R. de Ritis, A.A. Marino, \cqg {\bf
14}, 3243 (1997).
\bibitem{gottloeber} S. Gottl\"ober, H.-J. Schmidt, and A.A.
Starobinsky, \cqg {\bf 7}, 893 (1990).
\bibitem{ruzmaikin} T.V. Ruzmaikina and A.A. Ruzmaikin, {\it
JETP}, {\bf 30}, 372 (1970).
\bibitem{sixth} L. Amendola, A. Battaglia-Mayer, S. Capozziello,
S. Gottl\"ober, V. M\"uller, F. Occhionero and H.-J. Schmidt, \cqg
{\bf 10}, L43 (1993).
\bibitem{eight} A. Battaglia-Mayer and H.-J. Schmidt, \cqg
{\bf 10}, 2441 (1993).
\bibitem{schmidt1} H.-J. Schmidt, \cqg {\bf 7}, 1023 (1990).
\bibitem{aclo} L. Amendola, S.
Capozziello, M. Litterio, F. Occhionero, \pr {\bf  D 45}, 417
(1992).
\bibitem{lambdat} S. Capozziello, R. de Ritis, A.A. Marino, \grg {\bf
30}, 1247 (1998).
\bibitem{weinberg1} S. Weinberg, \rmp {\bf 61}, 1 (1989).
\bibitem{linde} A.D. Linde, \pl {\bf B 108}, 389 (1982).
\bibitem{hoyle} F. Hoyle and J.V. Narlikar, {Proc. R. Soc.A} {\bf
273}, 1 (1963).
\bibitem{snIa} J.L. Tonry, B.P. Schmidt, B. Barris {\it et al.}
\apj {\bf 594}, 1 (2003).
\bibitem{WMAP} http://map.gsfc.nasa.gov/
\bibitem{kenyon}I.R. Kenyon, {\it General Relativity}, Oxford
University Press, Oxford (1995).
\bibitem{cotsakis} S. Cotsakis and G. Flessas, \pl {\bf B 319}, 69
(1993).
\bibitem{waldPRD} R.M. Wald, \pr {D 28}, 2118 (1983).
\bibitem{hawking} S.W. Hawking and G.F.R. Ellis, {\it The Large-Scale Structure of
Space-Time}, Cambridge University Press, Cambridge (1973).
\bibitem{lambdastep} S. Capozziello, R. de Ritis, A.A. Marino, {\it Helv. Phys.
Acta} {\bf 69}, 241 (1996).
\bibitem{barth} N.H. Barth and Christensen, \pr {\bf 28}, 1876 (1983).
\bibitem{damour} T. Damour and G. Esposito-Farese, \cqg {\bf 9},
2093 (1994).
\bibitem{CCJ} C.G. Callan, S. Coleman, and R. Jackiw, {\it Ann.
Phys.} (N.Y.) {\bf 59}, 42 (1970).
\bibitem{penrose} R. Penrose, {\it Proc. R. Soc., Ser. A}, {\bf
284}, 159 (1965).
\bibitem{dickey} J.O. Dickey {\it et al.}, {\it Science} {\bf
265}, 482 (1982).
\bibitem{uzan} J. Ph. Uzan, \rmp {\bf 75}, 403 (2003).
\bibitem{damouresp} T.Damour and G. Esposito-Farese, \pr {\bf D
54}, 1474 (1996).
\bibitem{nmc} S. Capozziello, M. Demianski, R. de Ritis, and C.
Rubano, \pr {\bf D} {\bf 52}, 3288 (1995).
\bibitem{cqg} S. Capozziello and R. de Ritis, \pl {\bf 177 A} (1993) 1.\\
              S. Capozziello and R. de Ritis, \cqg {\bf 11} (1994) 107.
\bibitem{quartic}
S. Capozziello, R. de Ritis, C. Rubano, and P. Scudellaro, \ijmp
{\bf D 4}, 767 (1995).
\bibitem{nmdc} S. Capozziello and G. Lambiase, \grg {\bf 31}, 1005 (1999).
\bibitem{vafa} A.A. Tseytlin and C. Vafa, \np {\bf 372B} (1992) 443.
\bibitem{gasperini} M. Gasperini, {\it Proc. of the 2nd SIGRAV School on Astrophysics,
Cosmology and String Theory}, V. Gorini et al. eds. IOP Pub.,
Bristol 1999.
\bibitem{schwarz} M.B. Green, J.H. Schwarz, and E. Witten,
{\it Superstring theory}, Cambridge Univ. Press, Cambridge 1987.
\bibitem{tseytlin} A.A. Tseytlin and C. Vafa, \np {\bf 372B} (1992) 443.
\bibitem{caplamb} S. Capozziello and G. Lambiase, \ijmp {\bf D 12},
843 (2003).
\bibitem{giovannini} M. Gasperini, M. Giovannini, and G.
Veneziano, \prl {\bf 75} (1995) 3796.
\bibitem{kenmoku} M. Kenmoku, Y. Okamoto, and K. Shigemoto, \pr {\bf D
48} (1993) 578.
\bibitem{gibbons} G.W. Gibbons and B.F. Whiting, {\it Nature} {\bf 291} (1981) 636.
\bibitem{fischbach}
E. Fischbach, D. Sudarsky, A. Szafer, C. Talmadge, and S.H.
Aroson, Phys. Rev. Lett. {\bf 56}, 3 (1986) .
\bibitem{speake}
C.C. Speake and T.J. Quinn, Phys. Rev. Lett. {\bf 61}, 1340
(1988).
\bibitem{eckhardt1} D.H. Eckhardt, C. Jekeli, A.R. Lazarewicz, A.J.
Romaides, Phys. Rev. Lett. {\bf 60}, 2567 (1988).
\bibitem{fujii1} Y. Fujii, Phys. Lett.  {\bf B 202} 246, (1988).
\bibitem{mcgaugh} S.S. McGaugh, Ap. J. Lett. {\bf 541}, L33
(2000).
\bibitem{debernardis} P. de Bernardis, et al., Nature {\bf
404}, 955 (2000).
\bibitem{gasperini2} M. Gasperini, {\it Proc. of the 12th It. Conf. on General Relativity},
M. Bassan et al. eds., Worl Scientific, Singapore 1997.


\end{thebibliography}
\end{document}